\documentclass[
reprint,
superscriptaddress,
amsmath,
amssymb,
showkeys,
aip,
apl,
floatfix,
]{revtex4-2}

\usepackage{threeparttable}
\usepackage{setspace}
\usepackage{makecell}
\usepackage{amsmath}
\usepackage{graphicx}
\usepackage{dcolumn}
\usepackage{bm}
\usepackage{hyperref}
\usepackage[mathlines, pagewise]{lineno}
\usepackage{textcomp}
\usepackage{gensymb}
\usepackage{subfigure}
\usepackage{url}
\usepackage{xcolor}
\usepackage{soul}
\usepackage{hhline}
\usepackage[version=3]{mhchem}
\usepackage{multirow}
\usepackage[american]{babel}
\usepackage{hhline}
\usepackage{float}
\usepackage{textgreek}

\definecolor{red}{rgb}{0.6,0.1,0.1}\def\red{\color{red}}
\definecolor{blue}{rgb}{0.1,0.1,0.6}
\definecolor{green}{rgb}{0.1,0.6,0.1}

\begin{document}

\title{
Orientation-Dependent Atomic-Scale Mechanism of \texorpdfstring{\textbeta-\ce{Ga2O3}}{} Thin Film Epitaxial Growth
}

\author{Jun Zhang} 
\thanks{Jun Zhang and Junlei Zhao contribute to this work equally.}

\author{Junlei Zhao} 
\thanks{Jun Zhang and Junlei Zhao contribute to this work equally.}

\author{Junting Chen} 

\author{Mengyuan Hua} 
\email{huamy@sustech.edu.cn}
\affiliation{Department of Electrical and Electronic Engineering, Southern University of Science and Technology, Shenzhen 518055, China}


\begin{abstract}

$\beta$-\ce{Ga2O3} has gained intensive interests of research and application as an ultrawide bandgap semiconductor. 
Epitaxial growth technique of the $\beta$-\ce{Ga2O3} thin film possesses a fundamental and vital role in the \ce{Ga2O3}-based device fabrication. 
In this work, epitaxial growth mechanisms of $\beta$-\ce{Ga2O3} with four low Miller-index facets, namely (100), (010), (001), and ($\overline{2}01$), are systematically explored using large-scale machine-learning molecular dynamics simulations at the atomic scale.
The simulations reveal that the migration of the face-centered cubic stacking \ce{O} sublattice plays a predominant role in rationalizing the different growth mechanisms between (100)/(010)/(001) and ($\overline{2}01$) orientations.
The resultant complex combinations of the stacking faults and twin boundaries are carefully identified, and shows a good agreement with the experimental observation and \textit{ab initio} calculation.     
Our results provide useful insights into the gas-phase epitaxial growth of the $\beta$-\ce{Ga2O3} thin films and suggest possible ways to tailor its properties for specific applications.

\end{abstract}
\date{\today}

\maketitle


$\beta$-\ce{Ga2O3} emerges as an ultrawide bandgap semiconductor ($E_\mathrm{g}\simeq4.8$ eV) with high promises for next-generation power electronics~\cite{pearton2018a, green2022gallium, zhang2022ultra}, solar-blind ultraviolet photodetector~\cite{xu2019gallium, kalra2022road}, high-temperature gas sensing~\cite{lin2017uv, SFzhao2021two}, and radiation-resistant device~\cite{SFazarov2023universal}.
The unique combination of availability for high-quality $\beta$-\ce{Ga2O3} bulk substrate grown by melt-grown method~\cite{galazka2014on, galazka2022growth, galazka2022two} and widely tunable $n$-type conductivity makes it competitive to the successfully commercialized wide bandgap semiconductors \ce{SiC} and \ce{GaN} in terms of both manufacturing cost and device performance~\cite{tsao2018ultrawide, reese_how_2019, heinselman2022projected}. 

As the most stable polymorphic phase among the five experimentally known ones (designated as $\beta$, $\kappa$, $\alpha$, $\delta$, and $\gamma$)~\cite{yoshioka2007structures, mu2022phase}, $\beta$-\ce{Ga2O3} has a monoclinic crystal structure (space group, $C2/m$) with strong anisotropy and significant orientation-dependent properties~\cite{ueda1997anisotropy, jiang2018three, mu2020first}.
This low-symmetry structural nature of $\beta$-\ce{Ga2O3} results in easily cleavable (100) and ($\overline{2}01$) planes and non-cleavable (010) and (001) planes, and further leads to distinct differences in the growth process and rate along these planes~\cite{tsai2010beta, kaun2015beta}.  
As an critical technique in semiconductor industry, the homoepitaxial growth of $\beta$-\ce{Ga2O3} thin film undergoes intensive experimental progresses from different growth methods in parallel, including molecular beam epitaxy (MBE)~\cite{2013_sasaki_MBE, 2014_okumura_MBE, 2018_oshima_MBE, 2023_Azizie_S-MBE}, metal-organic chemical vapor deposition (MOCVD)~\cite{2022_MengLY_MOCVD, tang2022high}, atomic layer deposition (ALD)~\cite{2005_Shan_ALD}, and metal-oxide catalyzed epitaxy (MOCATAXY)~\cite{2020_mauze_MOCATAXY}. 
The structural properties of various imperfections (e.g., point defect, twin boundary (TB), and stacking fault (SF)), and their effects on the device performance are under careful experimental scrutiny~\cite{2014_wagner_homoepitaxial_stacking,2017_fiedler_influence_stackingfault, 2020_eisner_201_010, 2021_sdoeung_stacking_leakage}. 

However, these experimental studies were so far mainly complemented by expensive \textit{ab initio} calculations~\cite{bermudez2006the, peelaers2015brillouin, mu2020first, mu2022phase, wang2023first}. 
The vital information involving complex kinetic and thermodynamic evolution of the growth processes at the atomic level is yet missing. 
Therefore, in this work, we employ large-scale molecular dynamics (MD) simulation enabled by our recently developed machine-learning interatomic potential (ML-IP) of \ce{Ga2O3} system~\cite{SFzhao2023complex} to simulate the orientation-dependent dynamic processes of the initial deposition and subsequent annealing of the homoepitaxial $\beta$-\ce{Ga2O3} thin films. 
We emphasize the different mobilities of \ce{O} and \ce{Ga} sublattices seen in the different lattice orientations, and their interplay with the resultant structural imperfections of the thin film.
Moreover, we show that the present of the twin boundaries and stacking faults on ($\overline{2}01$) are profoundly predominated by the misaligned stacking ordering of the \ce{O} sublattice.

\begin{figure}[htb]
\centering
\includegraphics[width=8cm]{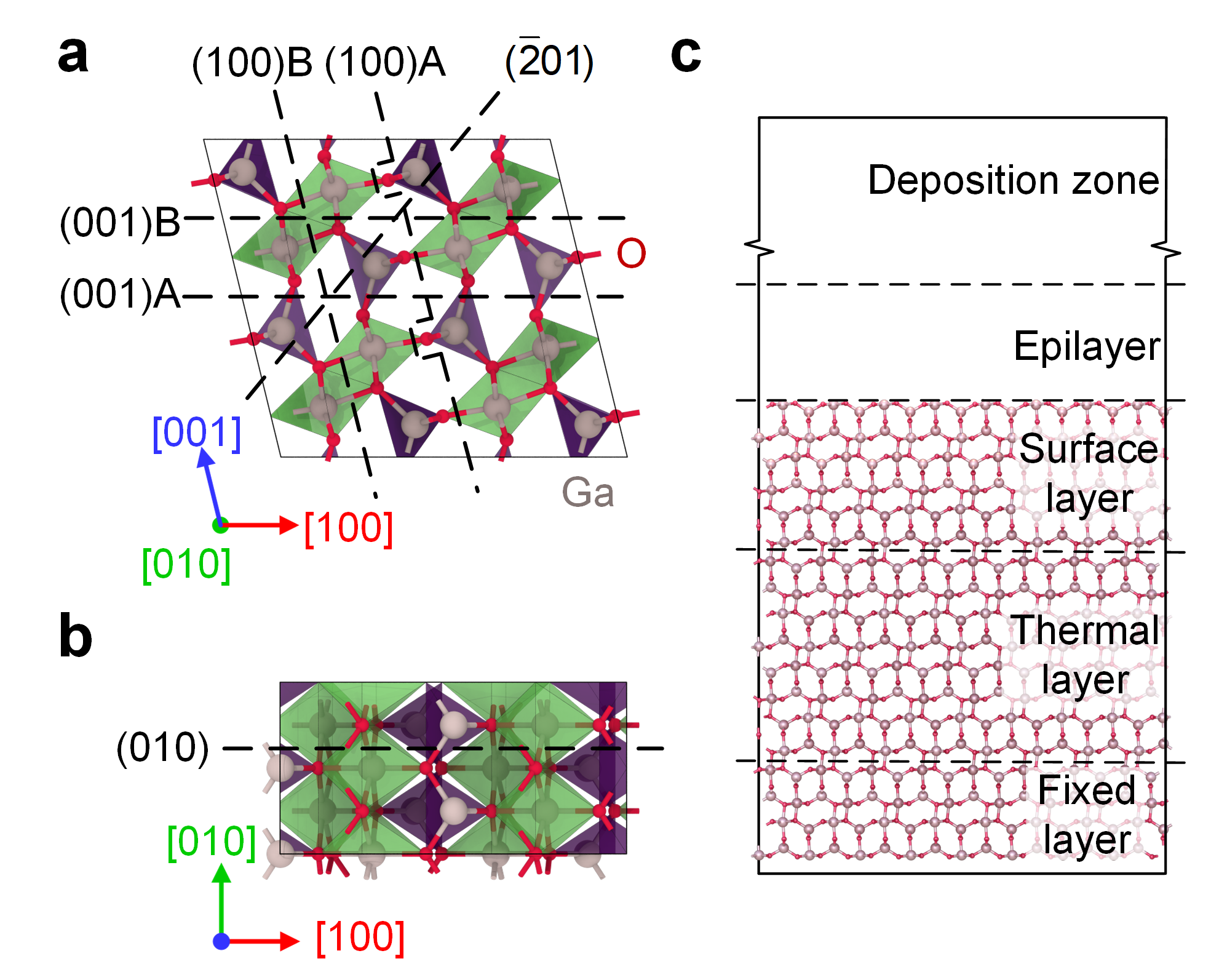}
\caption{Side view (a) and top view (b) of the 80-atom supercell of monoclinic $\beta$-\ce{Ga2O3}. The polyhedra with \ce{Ga} centers (in brown) and \ce{O} (in red) corners are color coded with the coordination numbers. 4-fold (tetrahedral) and 6-fold (octahedral) ones are in purple and green, respectively. The dashed lines indicate the cutting planes. (c) The exemplary simulation cell with (100) surface. The dashed lines separate the different simulation regions.}
\label{fig1}
\end{figure}

We choose four lattice orientations with low Miller indices, namely, (100), (010), (001), and ($\overline{2}01$), as shown in Figs.~\ref{fig1}a and \ref{fig1}b, where the dashed lines indicate the cutting planes in a $1\times2\times2$ 80-atom monoclinic supercell. 
For (100) and (001) orientations, two non-equivalent surface terminations are labelled as types A and B, adopted from the notation in Ref.~\citenum{mu2020first} where the surface energies of the corresponding surfaces are accurately calculated using \textit{ab initio} method. 
Here, we use the (100)B and (001)B surface terminations with the lower surface energies to construct the two initial slabs of (100) and (001) orientations, respectively.  
An exemplary initial setup of the (100) simulation cell is illustrated in Fig.~\ref{fig1}c. 
The dimensions of the cells are $\sim50\times\sim50\times\sim150$ \r A$^{3}$ with marginal differences in $x$ and $y$ side lengths to keep the periodical boundary conditions and are further scaled with the thermal expansion coefficients of the certain temperatures.

The initial numbers of atoms are kept in stoichiometric proportions ($\ce{Ga}:\ce{O}=2:3$), and are 12240, 11520, 11560, and 11880 for (100), (010), (001), and ($\overline{2}01$) orientations, respectively.
The total number of the deposited atoms are 4080, 4320, 4080 and 4320, which corresponds to the three full stoichiometric periodic layers.
All the simulation cells constitute of three groups of atoms depending on the initial positions:
(i) the atoms in the fixed layer are fixed to avoid the movement of the entire cell due to the momentum introduced by the incoming depositing atoms;
(ii) the atoms in the thermal layer are controlled by Nos{\'e}-Hoover thermostat; 
(iii) the atoms in surface layer, epilayer and deposition zone are allowed to moving freely following Newton's law. 
The isolated depositing atoms are created randomly in the deposition zone every 1200/800 MD step for \ce{Ga}/\ce{O}, respectively. 
The initial positions of the newly created atoms are ensured no overlapping ($> 5$ \r A) with the existing atoms. 
In this way, the depositing atoms can land on the surface following a stochastic process, with no direct artifact introduced by the thermostat, whereas the latent heat released during the deposition can be drained with the thermal layer through phonon transport. 
This quasi-canonical ensemble has been commonly used for the open-surface MD simulations~\cite{SFzhao2016formation, SFzhao2017formation}. 
The total simulation time is 2200 ps for epitaxial growth followed by 5000 ps annealing with 1 fs per MD step.
We simulate five temperatures as 300/600/900/1200/1500 K for all the four orientations, and three independent random simulations for each case, so $5\times4\times3=60$ simulations in total.
The MD simulations are conducted using LAMMPS package~\cite{lammps1995}, and the visualizations of the atomic structures are done using OVITO~\cite{2010_ovito_stukowski}. 

\begin{figure*}[htb!]
\centering
\includegraphics[width=15cm]{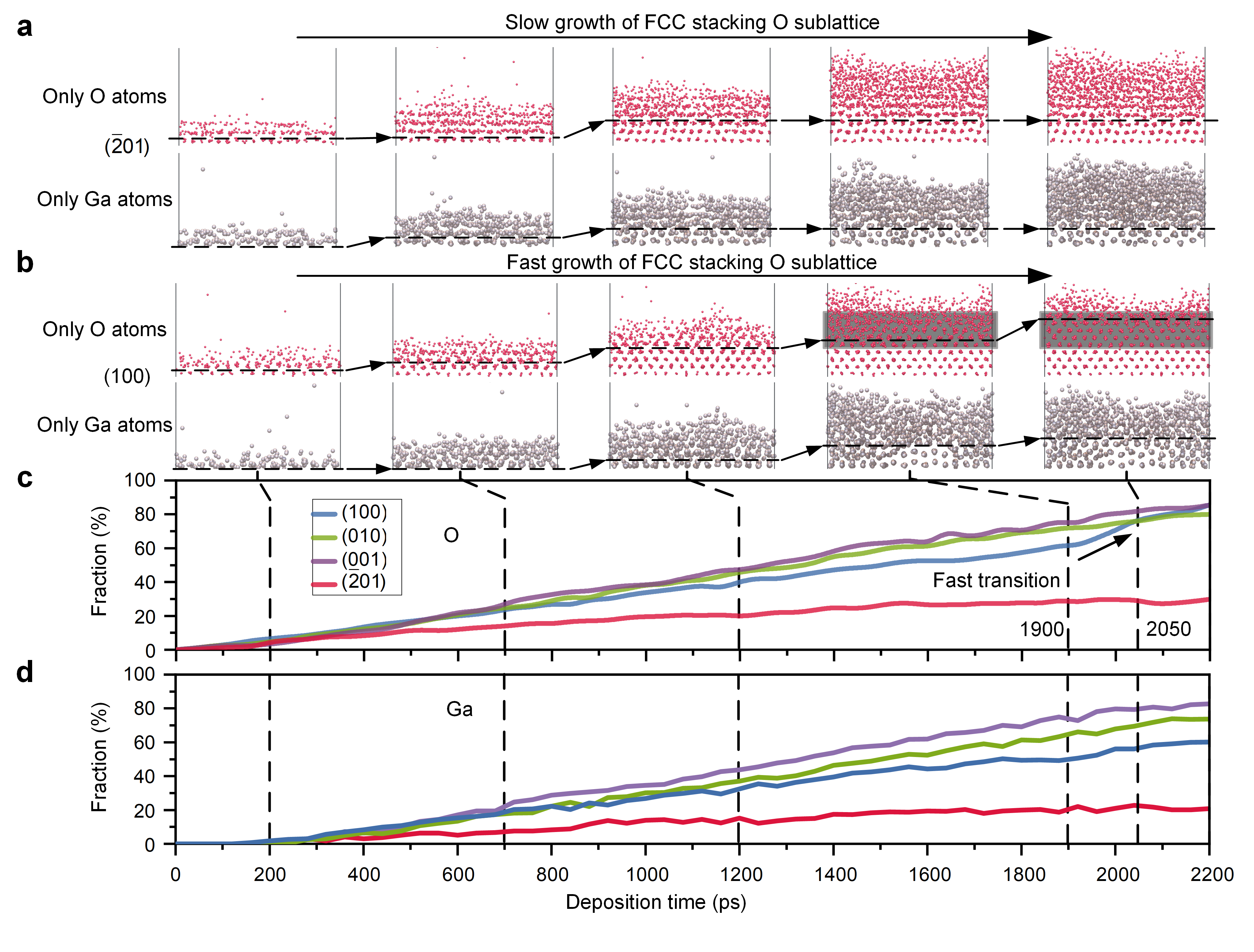}
\caption{The evolution of the \ce{O} (upper row) and \ce{Ga} (lower row) sublattices of the deposited epilayers on (a) ($\overline{2}01$) and (b) (100)B facets during growth process. 
The horizontal dashed lines distinguish the ordered/disordered atoms. The grey shadow regions in (b) emphasize the fast reordering transition of \ce{O} on the (100) facet in 150 ps. The evolution of the atomic factions of the ordered (c) \ce{O} and (d) \ce{Ga} atoms.}
\label{fig2}
\end{figure*}

Figs.~\ref{fig2}a and \ref{fig2}b show the representative deposition processes of $\beta$-\ce{Ga2O3} ($\overline{2}01$) and (100)B at 1200 K, respectively, with the \ce{O} and \ce{Ga} sublattices separately shown in the upper and lower rows. 
The ordered/disordered sublattice interfaces are marked by dashed lines, determined by the Wigner-Seitz (WS) analysis for point defects~\cite{2020W-zmethod}. This analysis can additionally serve to quantitatively determine the fraction of ordered atoms among the total number of deposited atoms, as shown in Figs~\ref{fig2}c and \ref{fig2}d.
Intriguingly, the growth process on the ($\overline{2}01$) facet exhibits a slow reordering of the FCC \ce{O} sublattices, compared with the other three facets. 
This difference can be clearly seen from the WS analysis in Figs~\ref{fig2}c and \ref{fig2}d as both of the fractions of the ordered \ce{O} and \ce{Ga} atoms are below 30 \% for the ($\overline{2}01$) facet, whereas that of the other three facets are around 70 \% for the as-grown thin films. 
This phenomenon can be elucidated with the orientations of the \ce{O} FCC sublattice.
Specifically, the ($\overline{2}01$) facets of the $\beta$-\ce{Ga2O3} corresponds to the (111) FCC \ce{O} facets. 
It is well-known that FCC (111) is a low-energy slip plane (with slip direction [110]), therefore, the reordering of the FCC \ce{O} sublattice can be hindered by the random mis-alignment in the (111) orientation.
We will discuss with this aspect when talking about the annealing process later on.

On the other hand, notably, a fast reordering transition of the \ce{O} FCC sublattice is seen in the (100) growth within 150 ps, labeled by the black arrow in Fig.~\ref{fig2}c and correspondingly shown as the \ce{O} atoms in the shadowed boxes in Fig.~\ref{fig2}b.
This process is similar to the fast rearrangement of the \ce{O} FCC sublattice seen at the surface-free interface of the liquid-solid phase transition~\cite{SFzhao2023complex}, indicating already a quasi-bulk behavior of the as-grown epilayer underneath the immediate surface of two-to-three atomic layers.
Furthermore, the temperature effect on the growth process has a similar trend for the four facets (see the supplementary material Fig. S2). 
The atomic factions of the ordered \ce{Ga}/\ce{O} atoms increase from 300 K to 1200 K, indicating a promoted reordering transition with the increased mobilities of atoms. 
However, at 1500 K, the growth of the thin film can lead to a partial melting process that leads to a more disordered lattice.
Therefore, we now further focus on the annealing process at 1200 K to have a significant recovery of lattice during the 5000-ps annealing process, and yet without potential surface meting effect (see the supplementary material Fig. S3).  

\begin{figure*}[htb!]
\centering
\includegraphics[width=15cm]{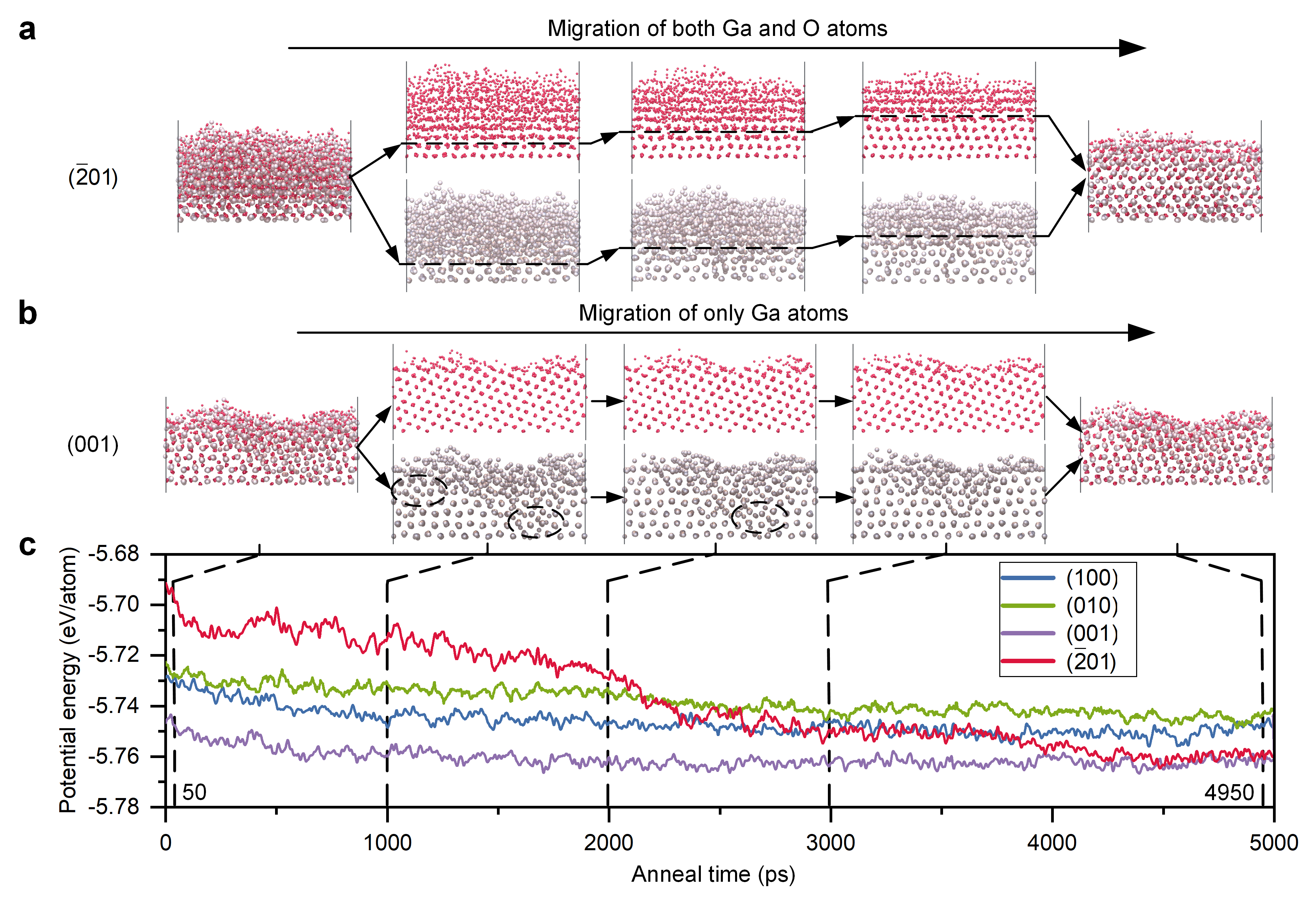}
\caption{Evolution of \ce{Ga} and \ce{O} sublattices of the epilayers on (a) ($\overline{2}01$) and (b) (001) facets during annealing process at 1200 K. \ce{Ga} is in brown and \ce{O} is in red.
The horizontal dashed lines distinguish the ordered/disordered atoms. 
The dashed circles in (b) highlight \ce{Ga} defect clusters. 
(c) Potential energy of the epilayers as a function of annealing time.}
\label{fig3}
\end{figure*}

As shown in Fig.~\ref{fig3}a, for the case of ($\overline{2}01$) orientation, annealing at 1200 K causes the further reordering (recovery) transition of the thin film, in both \ce{O} and \ce{Ga} sublattices. 
This is consistent with the significant decrease of the potential energy (0.07 eV per atom), as shown in Fig.~\ref{fig3}c. 
However, the WS analysis of point defects cannot catch this transition process because of the extended stacking faults. 
This point will be discussed in detail later (Fig~\ref{fig5}).
On the other hand, the recovery process in the (001) orientation is mainly involve migration of \ce{Ga} atoms, as can be seen as the vanished defect clusters marked by the dashed circles in Fig.~\ref{fig3}b.
We note that the (100) and (010) orientations follow the similar recovery of the \ce{Ga} sublattice as that of the (001) with small decreases of the potential energies (0.03 eV per atom). 
In fact, \ce{O} sublattice grown on these three facets have established a relatively completed FFC lattice already from the growth process (Fig~\ref{fig2}c), so the annealing mainly lead to the migration of the \ce{Ga} atoms gradually recovering to perfect $\beta$-phase.

\begin{figure}[htb!]
\centering
\includegraphics[width=8cm]{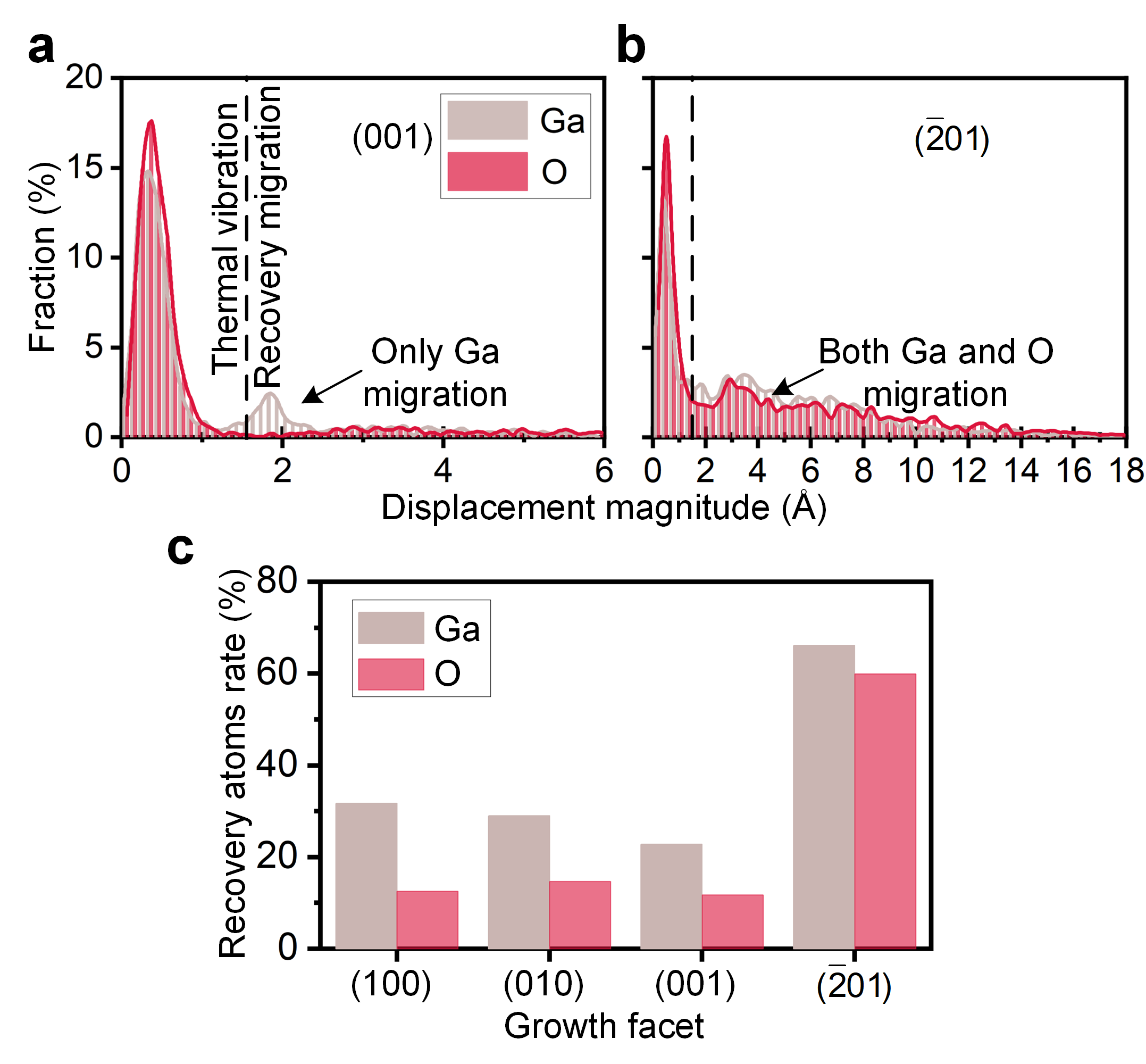}
\caption{Analysis of the atomic displacement magnitude during annealing process on (a) (001) and (b) ($\overline{2}01$). 1.6 \r A is the shortest \ce{Ga}-\ce{O} bond distance, which is set as the threshold to distinguish the thermal vibration and recovery migration. 
(c) The fraction of \ce{Ga} and \ce{O} atoms performing migration displacement for the four facets.
}
\label{fig4}
\end{figure}

\begin{figure*}[htb!]
\centering
\includegraphics[width=15cm]{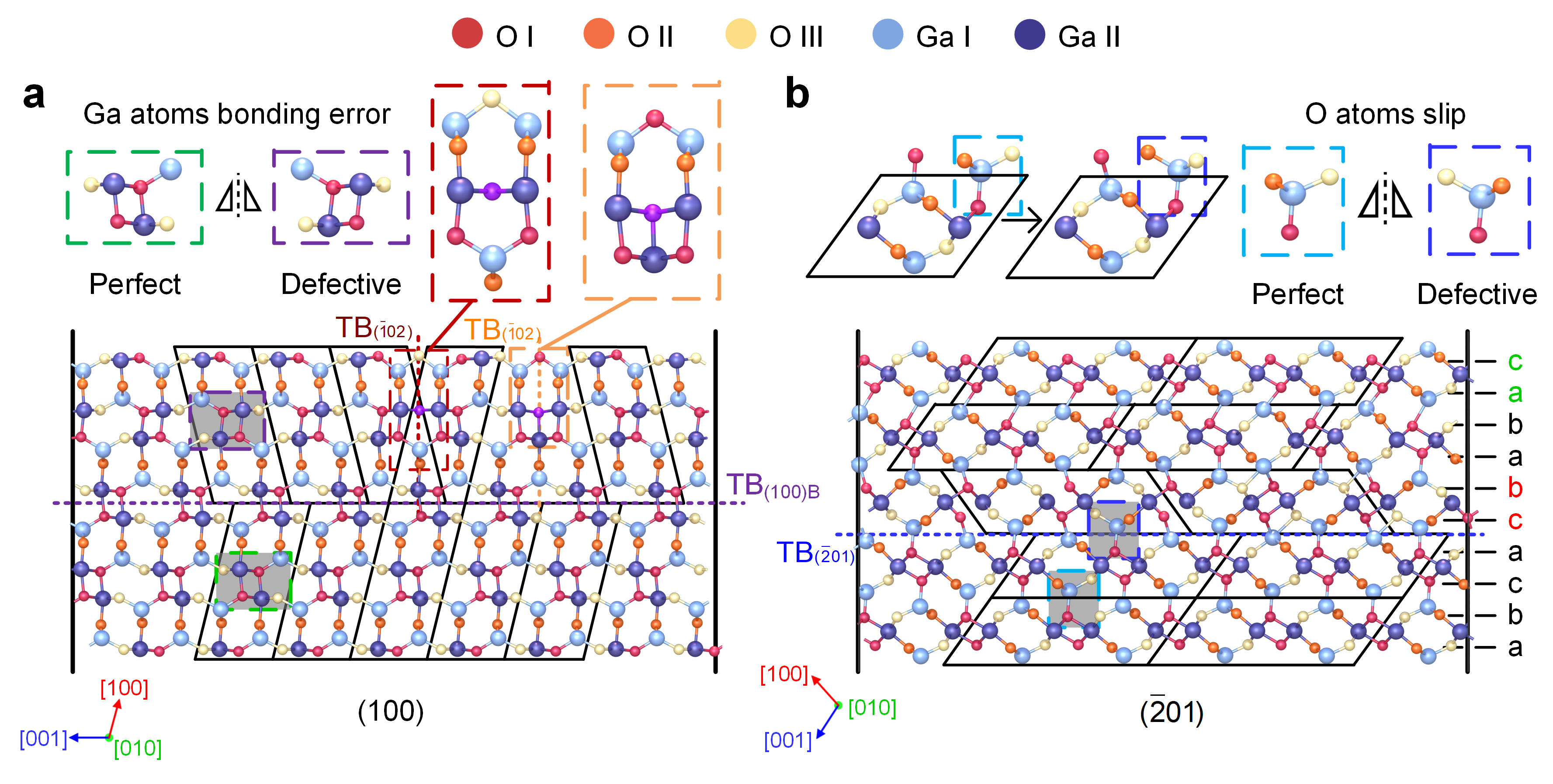}
\caption{Structures of the SF and TB identified on (a) (100) and (b) ($\overline{2}01$) planes of $\beta$-\ce{Ga2O3}. 
The dotted line indicate the twin boundary on corresponding plane, and unit cells are shown in solid case. 
The green and purple box in (a) represent the core of the unit cell on both sides of the TB.
The orange and red box display the enlarged models of atomic bonding at the TB on ($\overline{1}02$).
The light blue and dark blue dashed box in (b) represent the enlarged model at the perfect $\beta$-\ce{Ga2O3} boundary and TB on ($\overline{2}01$).
Letters on the right side denote the repetitive sequence of \ce{O} atoms, with letters in black displayed the normal structure of \ce{O} and letters in other colors indicated the various sacking fault. 
 }
\label{fig5}
\end{figure*}

To quantitatively elucidate the different recovery processes between the ($\overline{2}01$) and the other three facets, we plot the distributions of atomic displacements of the deposited atoms on (001) and ($\overline{2}01$), comparing between the snapshots at 0~ps and 5000~ps. 
As shown in Fig.~\ref{fig4}a and \ref{fig4}b, the major peaks below the displacement magnitude of 1.6 \r A (the length of the shortest \ce{Ga}-\ce{O} bond) correspond to the displacement introduced by the thermal vibration of the lattice,
whereas the distinct difference is revealed in the recovery migration range ($> 1.6$ \r A) of the distributions.
Unlike the (001) case where only clear \ce{Ga} migration peak is seen (Fig.~\ref{fig4}a), the ($\overline{2}01$) case has widely spread tails in both \ce{Ga} and \ce{O} distributions (Fig.~\ref{fig4}b), indicating an overall lattice rearrangement of the grown ($\overline{2}01$) thin film.  
A summary histogram plot in Fig.~\ref{fig4}c shows that more than 60 \% of the \ce{Ga}/\ce{O} atoms have gone through the lattice migration in the ($\overline{2}01$) thin film, much more than that in the other three facets.
See supplementary material {\red Fig. S4} for the distributions of all the four orientations.


The origins and effects of stacking faults and twins boundaries in the $\beta$-\ce{Ga2O3} epitaxial thin film are of tremendous interests for the experimental studies and applications for power device~\cite{2014_wagner_homoepitaxial_stacking, 2021_sdoeung_stacking_leakage, wang2023first}.
We now focus on the orientation-dependent formations of such defects in the annealed thin films. 
It has been discussed (Fig~\ref{fig2}a) that the (111) low-energy slip plane of the \ce{O} FCC sublattice has a predominant role in the formation of the planar defects, such as stacking fault and twin boundary seen in ($\overline{2}01$) facet. 
Moreover, we will show that with the low-symmetry \ce{Ga} sublattice, \ce{Ga}-related planar defects can form with the perfect \ce{O} FCC sublattice in the other three facts as well.

Here we analyze the representative defects seen in the MD simulations of the (100) and ($\overline{2}01$) facets. 
As shown in Fig~\ref{fig5}a, the \ce{O} sublattice follows the perfect $abc-abc$ stacking, whereas the \ce{Ga} sublattice forms three types of twin boundaries simultaneously. 
The (100) twin boundary is seen as a collection of the Ga atoms bonding errors where the 4-fold and 6-fold \ce{Ga} sites switch the stacking orders, as labeled by the dashed purple box. 
This lead to a mirrored twin lattice with a displacement of $1/4 \cdot \mathbf{c}$ in [001] direction, as experimentally studied in Refs~\citenum{2016_schewski_evolution_stacking_001, 2017_fiedler_influence_stackingfault}.
Meanwhile, two types of the ($\overline{1}$02) twin boundaries are seen by coalescing the mirrored domains with the original domains on (100) facets, as labeled by the dashed orange and red boxes. 
Purple \ce{O} atoms bonding only with the 6-fold \ce{Ga} atoms cannot be classified as the three intrinsic \ce{O} types, as typically \ce{O} atoms in $\beta$-\ce{Ga2O3} bond with 6- and 4-fold \ce{Ga} atoms simultaneously.
We note that these two ($\overline{1}$02) twin boundaries appear to be paired along the [001] direction owing to the lateral periodic conditions and surface termination, while they are paired along the [201] direction in a bulk twin boundary system~\cite{wang2023first}.

As shown in Fig~\ref{fig5}b, the different stacking faults of the \ce{O} FCC sublattice can lead to rather complex combination of the twin boundary and stacking fault. 
Starting form the perfect $abc$ stacking period, the next period shows an inverse order of $b$ and $c$ stacking layers (show in red letters), and results in a mirrored twin lattice with a displacement of $1/4 \cdot \mathbf{a} + 1/2 \cdot \mathbf{c}$. 
The stacking of the upper \ce{O} layers is largely affected by this twin boundary, and grows into hexagonal-like ($abab$ and/or $abac$) ordering (green and blue letters in Fig~\ref{fig5}b). 
These types of complex planar defects survived after 5000-ps annealing at 1200 K, indicating a much longer lifetime comparing to point defects. 
Intriguingly, on ($\overline{2}01$) facet, the local \ce{O}-\ce{Ga} bonding arrangements have no unidentified \ce{O} type (unlike the purple \ce{O} atoms in Fig~\ref{fig5}a), indicating that the relatively high mobilites of disordered \ce{O} atoms during the annealing process promote a short-range reordering regardless of the extended planar defects.           
In final remark, these atomic configurations of the stacking faults and twin boundaries shown in Fig~\ref{fig5}a and \ref{fig5}b are in exceptionally good agreement with the experimental and \textit{ab initio} studies~\cite{2014_wagner_homoepitaxial_stacking,2016_schewski_evolution_stacking_001, 2017_fiedler_influence_stackingfault, 2018_yamaguchi_stacking, 2020_eisner_201_010, wang2023first}. 

In an overview of the complete initial growth and subsequent annealing process, the non-slip orientations, such as (100), (010), and (001), can promote the qualities of as-grown epilayers with more ordered \ce{O} sublattice.
However, the limited mobility of the \ce{O} sublattice can hinder the further short-range lattice recovery during the annealing process.
The growth on the ($\overline{2}01$) facet can lead to significant planar defects involving the slip of the (111) \ce{O} planes, however, the annealing process may largely improve the local bonding intactness which can be favorable for device performance. 

In summary, leveraging the newly developed ML-MD simulations, the full dynamic processes of the orientation-dependent epitaxial growth are unveiled for the four low Miller-index facets of $\beta$-\ce{Ga2O3} system.
Our results show the migration of the FCC stacking \ce{O} sublattice plays a predominant role in explaining the different growth mechanisms between (100)/(010)/(001) and ($\overline{2}01$).
Through the atomic-scale analyses, \ce{Ga} defect states have the quantitative predominance on (100). (010), (001).
These results are in consistency with the experimental and \textit{ab initio} data, showing the good reliability of the ML-MD methods. 
Apart of the point defects, we show that the extended planar defects involving only \ce{Ga} or both \ce{Ga}\ce{O} sublattices are common imperfection configurations in the epitaxial growth of $\beta$-\ce{Ga2O3}.
The exact structures and dynamics of these systems are useful for the further improvements of the thin film quality and device performance.


See the supplementary material for the back-up simulations, more detailed analyses, and discussions.

This work is supported by Guangdong Basic and Applied Basic Research Foundation under Grant 2023A1515012048.
The computational resource is supported by the Center for Computational Science and Engineering at the Southern University of Science and Technology.

\section*{Author Declarations}

\subsection*{Conflict of Interest}

The authors have no conflicts to disclose.

\subsection*{Author Contributions}

\textbf{Jun Zhang:} Investigation (equal); Formal analysis (equal); Writing – original draft (lead).
\textbf{Junlei Zhao:} Conceptualization (lead); Investigation (lead); Methodology (lead); Formal analysis (lead); Writing – original draft (lead); Writing – review \& editing (equal); Resources (equal); Supervision (equal).
\textbf{Junting Chen:} Investigation (supporting); Formal analysis (supporting); Writing – review \& editing (equal).
\textbf{Mengyuan Hua:} Conceptualization (lead); Project administration (lead); Resources (equal); Supervision (equal); Writing – review \& editing (equal).

\section*{Data Availability}

The machine-learning potential parameter files used to run classical MD simulations are openly available at \url{https://doi.org/10.6084/m9.figshare.21731426}.
The data that support the findings of this study are available from the corresponding author upon reasonable request.

\bibliographystyle{apsrev4-2}
\bibliography{final}

\begin{thebibliography}{44}%
\makeatletter
\providecommand \@ifxundefined [1]{%
 \@ifx{#1\undefined}
}%
\providecommand \@ifnum [1]{%
 \ifnum #1\expandafter \@firstoftwo
 \else \expandafter \@secondoftwo
 \fi
}%
\providecommand \@ifx [1]{%
 \ifx #1\expandafter \@firstoftwo
 \else \expandafter \@secondoftwo
 \fi
}%
\providecommand \natexlab [1]{#1}%
\providecommand \enquote  [1]{``#1''}%
\providecommand \bibnamefont  [1]{#1}%
\providecommand \bibfnamefont [1]{#1}%
\providecommand \citenamefont [1]{#1}%
\providecommand \href@noop [0]{\@secondoftwo}%
\providecommand \href [0]{\begingroup \@sanitize@url \@href}%
\providecommand \@href[1]{\@@startlink{#1}\@@href}%
\providecommand \@@href[1]{\endgroup#1\@@endlink}%
\providecommand \@sanitize@url [0]{\catcode `\\12\catcode `\$12\catcode
  `\&12\catcode `\#12\catcode `\^12\catcode `\_12\catcode `\%12\relax}%
\providecommand \@@startlink[1]{}%
\providecommand \@@endlink[0]{}%
\providecommand \url  [0]{\begingroup\@sanitize@url \@url }%
\providecommand \@url [1]{\endgroup\@href {#1}{\urlprefix }}%
\providecommand \urlprefix  [0]{URL }%
\providecommand \Eprint [0]{\href }%
\providecommand \doibase [0]{https://doi.org/}%
\providecommand \selectlanguage [0]{\@gobble}%
\providecommand \bibinfo  [0]{\@secondoftwo}%
\providecommand \bibfield  [0]{\@secondoftwo}%
\providecommand \translation [1]{[#1]}%
\providecommand \BibitemOpen [0]{}%
\providecommand \bibitemStop [0]{}%
\providecommand \bibitemNoStop [0]{.\EOS\space}%
\providecommand \EOS [0]{\spacefactor3000\relax}%
\providecommand \BibitemShut  [1]{\csname bibitem#1\endcsname}%
\let\auto@bib@innerbib\@empty
\bibitem [{\citenamefont {Pearton}\ \emph {et~al.}(2018)\citenamefont
  {Pearton}, \citenamefont {Yang}, \citenamefont {Cary}, \citenamefont {Ren},
  \citenamefont {Kim}, \citenamefont {Tadjer},\ and\ \citenamefont
  {Mastro}}]{pearton2018a}%
  \BibitemOpen
  \bibfield  {author} {\bibinfo {author} {\bibfnamefont {S.~J.}\ \bibnamefont
  {Pearton}}, \bibinfo {author} {\bibfnamefont {J.}~\bibnamefont {Yang}},
  \bibinfo {author} {\bibfnamefont {P.~H.}\ \bibnamefont {Cary}}, \bibinfo
  {author} {\bibfnamefont {F.}~\bibnamefont {Ren}}, \bibinfo {author}
  {\bibfnamefont {J.~H.}\ \bibnamefont {Kim}}, \bibinfo {author} {\bibfnamefont
  {M.~J.}\ \bibnamefont {Tadjer}},\ and\ \bibinfo {author} {\bibfnamefont
  {M.~A.}\ \bibnamefont {Mastro}},\ }\href {https://doi.org/10.1063/1.5006941}
  {\bibfield  {journal} {\bibinfo  {journal} {Appl. Phys. Rev.}\ }\textbf
  {\bibinfo {volume} {5}},\ \bibinfo {pages} {011301} (\bibinfo {year}
  {2018})}\BibitemShut {NoStop}%
\bibitem [{\citenamefont {Green}\ \emph {et~al.}(2022)\citenamefont {Green},
  \citenamefont {Speck}, \citenamefont {Xing}, \citenamefont {Moens},
  \citenamefont {Allerstam}, \citenamefont {Gumaelius}, \citenamefont {Neyer},
  \citenamefont {Arias-PurdueAndrea}, \citenamefont {Mehrotra}, \citenamefont
  {Kuramata}, \citenamefont {Sasaki}, \citenamefont {Watanabe}, \citenamefont
  {Koshi}, \citenamefont {Blevins}, \citenamefont {Bierwagen}, \citenamefont
  {Krishnamoorthy}, \citenamefont {Leedy}, \citenamefont {Arehart},
  \citenamefont {Neal}, \citenamefont {Mou}, \citenamefont {Ringel},
  \citenamefont {Kumar}, \citenamefont {Sharma}, \citenamefont {Ghosh},
  \citenamefont {Singisetti}, \citenamefont {Li}, \citenamefont {Chabak},
  \citenamefont {Liddy}, \citenamefont {Islam}, \citenamefont {Rajan},
  \citenamefont {Graham}, \citenamefont {Choi}, \citenamefont {Cheng},\ and\
  \citenamefont {Higashiwaki}}]{green2022gallium}%
  \BibitemOpen
  \bibfield  {author} {\bibinfo {author} {\bibfnamefont {A.~J.}\ \bibnamefont
  {Green}}, \bibinfo {author} {\bibfnamefont {J.}~\bibnamefont {Speck}},
  \bibinfo {author} {\bibfnamefont {G.}~\bibnamefont {Xing}}, \bibinfo {author}
  {\bibfnamefont {P.}~\bibnamefont {Moens}}, \bibinfo {author} {\bibfnamefont
  {F.}~\bibnamefont {Allerstam}}, \bibinfo {author} {\bibfnamefont
  {K.}~\bibnamefont {Gumaelius}}, \bibinfo {author} {\bibfnamefont
  {T.}~\bibnamefont {Neyer}}, \bibinfo {author} {\bibnamefont
  {Arias-PurdueAndrea}}, \bibinfo {author} {\bibfnamefont {V.}~\bibnamefont
  {Mehrotra}}, \bibinfo {author} {\bibfnamefont {A.}~\bibnamefont {Kuramata}},
  \bibinfo {author} {\bibfnamefont {K.}~\bibnamefont {Sasaki}}, \bibinfo
  {author} {\bibfnamefont {S.}~\bibnamefont {Watanabe}}, \bibinfo {author}
  {\bibfnamefont {K.}~\bibnamefont {Koshi}}, \bibinfo {author} {\bibfnamefont
  {J.}~\bibnamefont {Blevins}}, \bibinfo {author} {\bibfnamefont
  {O.}~\bibnamefont {Bierwagen}}, \bibinfo {author} {\bibfnamefont
  {S.}~\bibnamefont {Krishnamoorthy}}, \bibinfo {author} {\bibfnamefont
  {K.}~\bibnamefont {Leedy}}, \bibinfo {author} {\bibfnamefont {A.~R.}\
  \bibnamefont {Arehart}}, \bibinfo {author} {\bibfnamefont {A.~T.}\
  \bibnamefont {Neal}}, \bibinfo {author} {\bibfnamefont {S.}~\bibnamefont
  {Mou}}, \bibinfo {author} {\bibfnamefont {S.~A.}\ \bibnamefont {Ringel}},
  \bibinfo {author} {\bibfnamefont {A.}~\bibnamefont {Kumar}}, \bibinfo
  {author} {\bibfnamefont {A.}~\bibnamefont {Sharma}}, \bibinfo {author}
  {\bibfnamefont {K.}~\bibnamefont {Ghosh}}, \bibinfo {author} {\bibfnamefont
  {U.}~\bibnamefont {Singisetti}}, \bibinfo {author} {\bibfnamefont
  {W.}~\bibnamefont {Li}}, \bibinfo {author} {\bibfnamefont {K.}~\bibnamefont
  {Chabak}}, \bibinfo {author} {\bibfnamefont {K.}~\bibnamefont {Liddy}},
  \bibinfo {author} {\bibfnamefont {A.}~\bibnamefont {Islam}}, \bibinfo
  {author} {\bibfnamefont {S.}~\bibnamefont {Rajan}}, \bibinfo {author}
  {\bibfnamefont {S.}~\bibnamefont {Graham}}, \bibinfo {author} {\bibfnamefont
  {S.}~\bibnamefont {Choi}}, \bibinfo {author} {\bibfnamefont {Z.}~\bibnamefont
  {Cheng}},\ and\ \bibinfo {author} {\bibfnamefont {M.}~\bibnamefont
  {Higashiwaki}},\ }\href {https://doi.org/10.1063/5.0060327} {\bibfield
  {journal} {\bibinfo  {journal} {APL Mater.}\ }\textbf {\bibinfo {volume}
  {10}},\ \bibinfo {pages} {029201} (\bibinfo {year} {2022})}\BibitemShut
  {NoStop}%
\bibitem [{\citenamefont {Zhang}\ \emph {et~al.}(2022)\citenamefont {Zhang},
  \citenamefont {Dong}, \citenamefont {Dang}, \citenamefont {Zhang},
  \citenamefont {Yan}, \citenamefont {Xiang}, \citenamefont {Su}, \citenamefont
  {Liu}, \citenamefont {Si}, \citenamefont {Gao}, \citenamefont {Kong},
  \citenamefont {Zhou},\ and\ \citenamefont {Hao}}]{zhang2022ultra}%
  \BibitemOpen
  \bibfield  {author} {\bibinfo {author} {\bibfnamefont {J.}~\bibnamefont
  {Zhang}}, \bibinfo {author} {\bibfnamefont {P.}~\bibnamefont {Dong}},
  \bibinfo {author} {\bibfnamefont {K.}~\bibnamefont {Dang}}, \bibinfo {author}
  {\bibfnamefont {Y.}~\bibnamefont {Zhang}}, \bibinfo {author} {\bibfnamefont
  {Q.}~\bibnamefont {Yan}}, \bibinfo {author} {\bibfnamefont {H.}~\bibnamefont
  {Xiang}}, \bibinfo {author} {\bibfnamefont {J.}~\bibnamefont {Su}}, \bibinfo
  {author} {\bibfnamefont {Z.}~\bibnamefont {Liu}}, \bibinfo {author}
  {\bibfnamefont {M.}~\bibnamefont {Si}}, \bibinfo {author} {\bibfnamefont
  {J.}~\bibnamefont {Gao}}, \bibinfo {author} {\bibfnamefont {M.}~\bibnamefont
  {Kong}}, \bibinfo {author} {\bibfnamefont {H.}~\bibnamefont {Zhou}},\ and\
  \bibinfo {author} {\bibfnamefont {Y.}~\bibnamefont {Hao}},\ }\href
  {https://doi.org/10.1038/s41467-022-31664-y} {\bibfield  {journal} {\bibinfo
  {journal} {Nat. Commun.}\ }\textbf {\bibinfo {volume} {13}},\ \bibinfo
  {pages} {3900} (\bibinfo {year} {2022})}\BibitemShut {NoStop}%
\bibitem [{\citenamefont {Xu}\ \emph {et~al.}(2019)\citenamefont {Xu},
  \citenamefont {Zheng},\ and\ \citenamefont {Huang}}]{xu2019gallium}%
  \BibitemOpen
  \bibfield  {author} {\bibinfo {author} {\bibfnamefont {J.}~\bibnamefont
  {Xu}}, \bibinfo {author} {\bibfnamefont {W.}~\bibnamefont {Zheng}},\ and\
  \bibinfo {author} {\bibfnamefont {F.}~\bibnamefont {Huang}},\ }\href
  {https://doi.org/10.1039/C9TC02055A} {\bibfield  {journal} {\bibinfo
  {journal} {J. Mater. Chem. C}\ }\textbf {\bibinfo {volume} {7}},\ \bibinfo
  {pages} {8753} (\bibinfo {year} {2019})}\BibitemShut {NoStop}%
\bibitem [{\citenamefont {Kalra}\ \emph {et~al.}(2022)\citenamefont {Kalra},
  \citenamefont {Muazzam}, \citenamefont {Muralidharan}, \citenamefont
  {Raghavan},\ and\ \citenamefont {Nath}}]{kalra2022road}%
  \BibitemOpen
  \bibfield  {author} {\bibinfo {author} {\bibfnamefont {A.}~\bibnamefont
  {Kalra}}, \bibinfo {author} {\bibfnamefont {U.~U.}\ \bibnamefont {Muazzam}},
  \bibinfo {author} {\bibfnamefont {R.}~\bibnamefont {Muralidharan}}, \bibinfo
  {author} {\bibfnamefont {S.}~\bibnamefont {Raghavan}},\ and\ \bibinfo
  {author} {\bibfnamefont {D.~N.}\ \bibnamefont {Nath}},\ }\href
  {https://doi.org/10.1063/5.0082348} {\bibfield  {journal} {\bibinfo
  {journal} {J. Appl. Phys.}\ }\textbf {\bibinfo {volume} {131}},\ \bibinfo
  {pages} {150901} (\bibinfo {year} {2022})}\BibitemShut {NoStop}%
\bibitem [{\citenamefont {Lin}\ \emph {et~al.}(2017)\citenamefont {Lin},
  \citenamefont {Gao},\ and\ \citenamefont {Gao}}]{lin2017uv}%
  \BibitemOpen
  \bibfield  {author} {\bibinfo {author} {\bibfnamefont {H.-J.}\ \bibnamefont
  {Lin}}, \bibinfo {author} {\bibfnamefont {H.}~\bibnamefont {Gao}},\ and\
  \bibinfo {author} {\bibfnamefont {P.-X.}\ \bibnamefont {Gao}},\ }\href
  {https://doi.org/10.1063/1.4974213} {\bibfield  {journal} {\bibinfo
  {journal} {Appl. Phys. Lett.}\ }\textbf {\bibinfo {volume} {110}},\ \bibinfo
  {pages} {043101} (\bibinfo {year} {2017})}\BibitemShut {NoStop}%
\bibitem [{\citenamefont {Zhao}\ \emph {et~al.}(2021)\citenamefont {Zhao},
  \citenamefont {Huang}, \citenamefont {Yin}, \citenamefont {Liao},
  \citenamefont {Mo}, \citenamefont {Qian}, \citenamefont {Guo}, \citenamefont
  {Chen}, \citenamefont {Zhang},\ and\ \citenamefont {Hua}}]{SFzhao2021two}%
  \BibitemOpen
  \bibfield  {author} {\bibinfo {author} {\bibfnamefont {J.}~\bibnamefont
  {Zhao}}, \bibinfo {author} {\bibfnamefont {X.}~\bibnamefont {Huang}},
  \bibinfo {author} {\bibfnamefont {Y.}~\bibnamefont {Yin}}, \bibinfo {author}
  {\bibfnamefont {Y.}~\bibnamefont {Liao}}, \bibinfo {author} {\bibfnamefont
  {H.}~\bibnamefont {Mo}}, \bibinfo {author} {\bibfnamefont {Q.}~\bibnamefont
  {Qian}}, \bibinfo {author} {\bibfnamefont {Y.}~\bibnamefont {Guo}}, \bibinfo
  {author} {\bibfnamefont {X.}~\bibnamefont {Chen}}, \bibinfo {author}
  {\bibfnamefont {Z.}~\bibnamefont {Zhang}},\ and\ \bibinfo {author}
  {\bibfnamefont {M.}~\bibnamefont {Hua}},\ }\href
  {https://doi.org/10.1021/acs.jpclett.1c01393} {\bibfield  {journal} {\bibinfo
   {journal} {J. Phys. Chem. Lett.}\ }\textbf {\bibinfo {volume} {12}},\
  \bibinfo {pages} {5813} (\bibinfo {year} {2021})}\BibitemShut {NoStop}%
\bibitem [{\citenamefont {Azarov}\ \emph {et~al.}(2023)\citenamefont {Azarov},
  \citenamefont {Fern{\'a}ndez}, \citenamefont {Zhao}, \citenamefont
  {Djurabekova}, \citenamefont {He}, \citenamefont {He}, \citenamefont {Prytz},
  \citenamefont {Vines}, \citenamefont {Bektas}, \citenamefont {Chekhonin},
  \citenamefont {Klingner}, \citenamefont {Hlawacek},\ and\ \citenamefont
  {Kuznetsov}}]{SFazarov2023universal}%
  \BibitemOpen
  \bibfield  {author} {\bibinfo {author} {\bibfnamefont {A.}~\bibnamefont
  {Azarov}}, \bibinfo {author} {\bibfnamefont {J.~G.}\ \bibnamefont
  {Fern{\'a}ndez}}, \bibinfo {author} {\bibfnamefont {J.}~\bibnamefont {Zhao}},
  \bibinfo {author} {\bibfnamefont {F.}~\bibnamefont {Djurabekova}}, \bibinfo
  {author} {\bibfnamefont {H.}~\bibnamefont {He}}, \bibinfo {author}
  {\bibfnamefont {R.}~\bibnamefont {He}}, \bibinfo {author} {\bibfnamefont
  {{\O}.}~\bibnamefont {Prytz}}, \bibinfo {author} {\bibfnamefont
  {L.}~\bibnamefont {Vines}}, \bibinfo {author} {\bibfnamefont
  {U.}~\bibnamefont {Bektas}}, \bibinfo {author} {\bibfnamefont
  {P.}~\bibnamefont {Chekhonin}}, \bibinfo {author} {\bibfnamefont
  {N.}~\bibnamefont {Klingner}}, \bibinfo {author} {\bibfnamefont
  {G.}~\bibnamefont {Hlawacek}},\ and\ \bibinfo {author} {\bibfnamefont
  {A.}~\bibnamefont {Kuznetsov}},\ }\href
  {https://doi.org/10.1038/s41467-023-40588-0} {\bibfield  {journal} {\bibinfo
  {journal} {Nat. Commun.}\ }\textbf {\bibinfo {volume} {14}},\ \bibinfo
  {pages} {4855} (\bibinfo {year} {2023})}\BibitemShut {NoStop}%
\bibitem [{\citenamefont {Galazka}\ \emph {et~al.}(2014)\citenamefont
  {Galazka}, \citenamefont {Irmscher}, \citenamefont {Uecker}, \citenamefont
  {Bertram}, \citenamefont {Pietsch}, \citenamefont {Kwasniewski},
  \citenamefont {Naumann}, \citenamefont {Schulz}, \citenamefont {Schewski},
  \citenamefont {Klimm},\ and\ \citenamefont {Bickermann}}]{galazka2014on}%
  \BibitemOpen
  \bibfield  {author} {\bibinfo {author} {\bibfnamefont {Z.}~\bibnamefont
  {Galazka}}, \bibinfo {author} {\bibfnamefont {K.}~\bibnamefont {Irmscher}},
  \bibinfo {author} {\bibfnamefont {R.}~\bibnamefont {Uecker}}, \bibinfo
  {author} {\bibfnamefont {R.}~\bibnamefont {Bertram}}, \bibinfo {author}
  {\bibfnamefont {M.}~\bibnamefont {Pietsch}}, \bibinfo {author} {\bibfnamefont
  {A.}~\bibnamefont {Kwasniewski}}, \bibinfo {author} {\bibfnamefont
  {M.}~\bibnamefont {Naumann}}, \bibinfo {author} {\bibfnamefont
  {T.}~\bibnamefont {Schulz}}, \bibinfo {author} {\bibfnamefont
  {R.}~\bibnamefont {Schewski}}, \bibinfo {author} {\bibfnamefont
  {D.}~\bibnamefont {Klimm}},\ and\ \bibinfo {author} {\bibfnamefont
  {M.}~\bibnamefont {Bickermann}},\ }\href
  {https://doi.org/10.1016/j.jcrysgro.2014.07.021} {\bibfield  {journal}
  {\bibinfo  {journal} {J. Cryst. Growth}\ }\textbf {\bibinfo {volume} {404}},\
  \bibinfo {pages} {184} (\bibinfo {year} {2014})}\BibitemShut {NoStop}%
\bibitem [{\citenamefont {Galazka}(2022)}]{galazka2022growth}%
  \BibitemOpen
  \bibfield  {author} {\bibinfo {author} {\bibfnamefont {Z.}~\bibnamefont
  {Galazka}},\ }\href {https://doi.org/10.1063/5.0076962} {\bibfield  {journal}
  {\bibinfo  {journal} {J. Appl. Phys.}\ }\textbf {\bibinfo {volume} {131}},\
  \bibinfo {pages} {031103} (\bibinfo {year} {2022})}\BibitemShut {NoStop}%
\bibitem [{\citenamefont {Galazka}\ \emph {et~al.}(2022)\citenamefont
  {Galazka}, \citenamefont {Ganschow}, \citenamefont {Seyidov}, \citenamefont
  {Irmscher}, \citenamefont {Pietsch}, \citenamefont {Chou}, \citenamefont
  {Bin~Anooz}, \citenamefont {Grueneberg}, \citenamefont {Popp}, \citenamefont
  {Dittmar}, \citenamefont {Kwasniewski}, \citenamefont {Suendermann},
  \citenamefont {Klimm}, \citenamefont {Straubinger}, \citenamefont
  {Schroeder},\ and\ \citenamefont {Bickermann}}]{galazka2022two}%
  \BibitemOpen
  \bibfield  {author} {\bibinfo {author} {\bibfnamefont {Z.}~\bibnamefont
  {Galazka}}, \bibinfo {author} {\bibfnamefont {S.}~\bibnamefont {Ganschow}},
  \bibinfo {author} {\bibfnamefont {P.}~\bibnamefont {Seyidov}}, \bibinfo
  {author} {\bibfnamefont {K.}~\bibnamefont {Irmscher}}, \bibinfo {author}
  {\bibfnamefont {M.}~\bibnamefont {Pietsch}}, \bibinfo {author} {\bibfnamefont
  {T.-S.}\ \bibnamefont {Chou}}, \bibinfo {author} {\bibfnamefont
  {S.}~\bibnamefont {Bin~Anooz}}, \bibinfo {author} {\bibfnamefont
  {R.}~\bibnamefont {Grueneberg}}, \bibinfo {author} {\bibfnamefont
  {A.}~\bibnamefont {Popp}}, \bibinfo {author} {\bibfnamefont {A.}~\bibnamefont
  {Dittmar}}, \bibinfo {author} {\bibfnamefont {A.}~\bibnamefont
  {Kwasniewski}}, \bibinfo {author} {\bibfnamefont {M.}~\bibnamefont
  {Suendermann}}, \bibinfo {author} {\bibfnamefont {D.}~\bibnamefont {Klimm}},
  \bibinfo {author} {\bibfnamefont {T.}~\bibnamefont {Straubinger}}, \bibinfo
  {author} {\bibfnamefont {T.}~\bibnamefont {Schroeder}},\ and\ \bibinfo
  {author} {\bibfnamefont {M.}~\bibnamefont {Bickermann}},\ }\href
  {https://doi.org/10.1063/5.0086996} {\bibfield  {journal} {\bibinfo
  {journal} {Appl. Phys. Lett.}\ }\textbf {\bibinfo {volume} {120}},\ \bibinfo
  {pages} {152101} (\bibinfo {year} {2022})}\BibitemShut {NoStop}%
\bibitem [{\citenamefont {Tsao}\ \emph {et~al.}(2018)\citenamefont {Tsao},
  \citenamefont {Chowdhury}, \citenamefont {Hollis}, \citenamefont {Jena},
  \citenamefont {Johnson}, \citenamefont {Jones}, \citenamefont {Kaplar},
  \citenamefont {Rajan}, \citenamefont {Van~de Walle}, \citenamefont
  {Bellotti}, \citenamefont {Chua}, \citenamefont {Collazo}, \citenamefont
  {Coltrin}, \citenamefont {Cooper}, \citenamefont {Evans}, \citenamefont
  {Graham}, \citenamefont {Grotjohn}, \citenamefont {Heller}, \citenamefont
  {Higashiwaki}, \citenamefont {Islam}, \citenamefont {Juodawlkis},
  \citenamefont {Khan}, \citenamefont {Koehler}, \citenamefont {Leach},
  \citenamefont {Mishra}, \citenamefont {Nemanich}, \citenamefont
  {Pilawa-Podgurski}, \citenamefont {Shealy}, \citenamefont {Sitar},
  \citenamefont {Tadjer}, \citenamefont {Witulski}, \citenamefont {Wraback},\
  and\ \citenamefont {Simmons}}]{tsao2018ultrawide}%
  \BibitemOpen
  \bibfield  {author} {\bibinfo {author} {\bibfnamefont {J.~Y.}\ \bibnamefont
  {Tsao}}, \bibinfo {author} {\bibfnamefont {S.}~\bibnamefont {Chowdhury}},
  \bibinfo {author} {\bibfnamefont {M.~A.}\ \bibnamefont {Hollis}}, \bibinfo
  {author} {\bibfnamefont {D.}~\bibnamefont {Jena}}, \bibinfo {author}
  {\bibfnamefont {N.~M.}\ \bibnamefont {Johnson}}, \bibinfo {author}
  {\bibfnamefont {K.~A.}\ \bibnamefont {Jones}}, \bibinfo {author}
  {\bibfnamefont {R.~J.}\ \bibnamefont {Kaplar}}, \bibinfo {author}
  {\bibfnamefont {S.}~\bibnamefont {Rajan}}, \bibinfo {author} {\bibfnamefont
  {C.~G.}\ \bibnamefont {Van~de Walle}}, \bibinfo {author} {\bibfnamefont
  {E.}~\bibnamefont {Bellotti}}, \bibinfo {author} {\bibfnamefont {C.~L.}\
  \bibnamefont {Chua}}, \bibinfo {author} {\bibfnamefont {R.}~\bibnamefont
  {Collazo}}, \bibinfo {author} {\bibfnamefont {M.~E.}\ \bibnamefont
  {Coltrin}}, \bibinfo {author} {\bibfnamefont {J.~A.}\ \bibnamefont {Cooper}},
  \bibinfo {author} {\bibfnamefont {K.~R.}\ \bibnamefont {Evans}}, \bibinfo
  {author} {\bibfnamefont {S.}~\bibnamefont {Graham}}, \bibinfo {author}
  {\bibfnamefont {T.~A.}\ \bibnamefont {Grotjohn}}, \bibinfo {author}
  {\bibfnamefont {E.~R.}\ \bibnamefont {Heller}}, \bibinfo {author}
  {\bibfnamefont {M.}~\bibnamefont {Higashiwaki}}, \bibinfo {author}
  {\bibfnamefont {M.~S.}\ \bibnamefont {Islam}}, \bibinfo {author}
  {\bibfnamefont {P.~W.}\ \bibnamefont {Juodawlkis}}, \bibinfo {author}
  {\bibfnamefont {M.~A.}\ \bibnamefont {Khan}}, \bibinfo {author}
  {\bibfnamefont {A.~D.}\ \bibnamefont {Koehler}}, \bibinfo {author}
  {\bibfnamefont {J.~H.}\ \bibnamefont {Leach}}, \bibinfo {author}
  {\bibfnamefont {U.~K.}\ \bibnamefont {Mishra}}, \bibinfo {author}
  {\bibfnamefont {R.~J.}\ \bibnamefont {Nemanich}}, \bibinfo {author}
  {\bibfnamefont {R.~C.~N.}\ \bibnamefont {Pilawa-Podgurski}}, \bibinfo
  {author} {\bibfnamefont {J.~B.}\ \bibnamefont {Shealy}}, \bibinfo {author}
  {\bibfnamefont {Z.}~\bibnamefont {Sitar}}, \bibinfo {author} {\bibfnamefont
  {M.~J.}\ \bibnamefont {Tadjer}}, \bibinfo {author} {\bibfnamefont {A.~F.}\
  \bibnamefont {Witulski}}, \bibinfo {author} {\bibfnamefont {M.}~\bibnamefont
  {Wraback}},\ and\ \bibinfo {author} {\bibfnamefont {J.~A.}\ \bibnamefont
  {Simmons}},\ }\href {https://doi.org/10.1002/aelm.201600501} {\bibfield
  {journal} {\bibinfo  {journal} {Adv. Electron. Mater.}\ }\textbf {\bibinfo
  {volume} {4}},\ \bibinfo {pages} {1600501} (\bibinfo {year}
  {2018})}\BibitemShut {NoStop}%
\bibitem [{\citenamefont {Reese}\ \emph {et~al.}(2019)\citenamefont {Reese},
  \citenamefont {Remo}, \citenamefont {Green},\ and\ \citenamefont
  {Zakutayev}}]{reese_how_2019}%
  \BibitemOpen
  \bibfield  {author} {\bibinfo {author} {\bibfnamefont {S.~B.}\ \bibnamefont
  {Reese}}, \bibinfo {author} {\bibfnamefont {T.}~\bibnamefont {Remo}},
  \bibinfo {author} {\bibfnamefont {J.}~\bibnamefont {Green}},\ and\ \bibinfo
  {author} {\bibfnamefont {A.}~\bibnamefont {Zakutayev}},\ }\href
  {https://doi.org/https://doi.org/10.1016/j.joule.2019.01.011} {\bibfield
  {journal} {\bibinfo  {journal} {Joule}\ }\textbf {\bibinfo {volume} {3}},\
  \bibinfo {pages} {903} (\bibinfo {year} {2019})}\BibitemShut {NoStop}%
\bibitem [{\citenamefont {Heinselman}\ \emph {et~al.}(2022)\citenamefont
  {Heinselman}, \citenamefont {Haven}, \citenamefont {Zakutayev},\ and\
  \citenamefont {Reese}}]{heinselman2022projected}%
  \BibitemOpen
  \bibfield  {author} {\bibinfo {author} {\bibfnamefont {K.~N.}\ \bibnamefont
  {Heinselman}}, \bibinfo {author} {\bibfnamefont {D.}~\bibnamefont {Haven}},
  \bibinfo {author} {\bibfnamefont {A.}~\bibnamefont {Zakutayev}},\ and\
  \bibinfo {author} {\bibfnamefont {S.~B.}\ \bibnamefont {Reese}},\ }\href
  {https://doi.org/10.1021/acs.cgd.2c00340} {\bibfield  {journal} {\bibinfo
  {journal} {Cryst. Growth Des.}\ }\textbf {\bibinfo {volume} {22}},\ \bibinfo
  {pages} {4854} (\bibinfo {year} {2022})}\BibitemShut {NoStop}%
\bibitem [{\citenamefont {Yoshioka}\ \emph {et~al.}(2007)\citenamefont
  {Yoshioka}, \citenamefont {Hayashi}, \citenamefont {Kuwabara}, \citenamefont
  {Oba}, \citenamefont {Matsunaga},\ and\ \citenamefont
  {Tanaka}}]{yoshioka2007structures}%
  \BibitemOpen
  \bibfield  {author} {\bibinfo {author} {\bibfnamefont {S.}~\bibnamefont
  {Yoshioka}}, \bibinfo {author} {\bibfnamefont {H.}~\bibnamefont {Hayashi}},
  \bibinfo {author} {\bibfnamefont {A.}~\bibnamefont {Kuwabara}}, \bibinfo
  {author} {\bibfnamefont {F.}~\bibnamefont {Oba}}, \bibinfo {author}
  {\bibfnamefont {K.}~\bibnamefont {Matsunaga}},\ and\ \bibinfo {author}
  {\bibfnamefont {I.}~\bibnamefont {Tanaka}},\ }\href
  {https://doi.org/10.1088/0953-8984/19/34/346211} {\bibfield  {journal}
  {\bibinfo  {journal} {J. Phys.: Condens. Matter}\ }\textbf {\bibinfo {volume}
  {19}},\ \bibinfo {pages} {346211} (\bibinfo {year} {2007})}\BibitemShut
  {NoStop}%
\bibitem [{\citenamefont {Mu}\ and\ \citenamefont {Van~de
  Walle}(2022)}]{mu2022phase}%
  \BibitemOpen
  \bibfield  {author} {\bibinfo {author} {\bibfnamefont {S.}~\bibnamefont
  {Mu}}\ and\ \bibinfo {author} {\bibfnamefont {C.~G.}\ \bibnamefont {Van~de
  Walle}},\ }\href {https://doi.org/10.1103/PhysRevMaterials.6.104601}
  {\bibfield  {journal} {\bibinfo  {journal} {Phys. Rev. Mater.}\ }\textbf
  {\bibinfo {volume} {6}},\ \bibinfo {pages} {104601} (\bibinfo {year}
  {2022})}\BibitemShut {NoStop}%
\bibitem [{\citenamefont {Ueda}\ \emph {et~al.}(1997)\citenamefont {Ueda},
  \citenamefont {Hosono}, \citenamefont {Waseda},\ and\ \citenamefont
  {Kawazoe}}]{ueda1997anisotropy}%
  \BibitemOpen
  \bibfield  {author} {\bibinfo {author} {\bibfnamefont {N.}~\bibnamefont
  {Ueda}}, \bibinfo {author} {\bibfnamefont {H.}~\bibnamefont {Hosono}},
  \bibinfo {author} {\bibfnamefont {R.}~\bibnamefont {Waseda}},\ and\ \bibinfo
  {author} {\bibfnamefont {H.}~\bibnamefont {Kawazoe}},\ }\href
  {https://doi.org/10.1063/1.119693} {\bibfield  {journal} {\bibinfo  {journal}
  {Appl. Phys. Lett.}\ }\textbf {\bibinfo {volume} {71}},\ \bibinfo {pages}
  {933} (\bibinfo {year} {1997})}\BibitemShut {NoStop}%
\bibitem [{\citenamefont {Jiang}\ \emph {et~al.}(2018)\citenamefont {Jiang},
  \citenamefont {Qian}, \citenamefont {Li},\ and\ \citenamefont
  {Yang}}]{jiang2018three}%
  \BibitemOpen
  \bibfield  {author} {\bibinfo {author} {\bibfnamefont {P.}~\bibnamefont
  {Jiang}}, \bibinfo {author} {\bibfnamefont {X.}~\bibnamefont {Qian}},
  \bibinfo {author} {\bibfnamefont {X.}~\bibnamefont {Li}},\ and\ \bibinfo
  {author} {\bibfnamefont {R.}~\bibnamefont {Yang}},\ }\href
  {https://doi.org/10.1063/1.5054573} {\bibfield  {journal} {\bibinfo
  {journal} {Appl. Phys. Lett.}\ }\textbf {\bibinfo {volume} {113}},\ \bibinfo
  {pages} {232105} (\bibinfo {year} {2018})}\BibitemShut {NoStop}%
\bibitem [{\citenamefont {Mu}\ \emph {et~al.}(2020)\citenamefont {Mu},
  \citenamefont {Wang}, \citenamefont {Peelaers},\ and\ \citenamefont {Van~de
  Walle}}]{mu2020first}%
  \BibitemOpen
  \bibfield  {author} {\bibinfo {author} {\bibfnamefont {S.}~\bibnamefont
  {Mu}}, \bibinfo {author} {\bibfnamefont {M.}~\bibnamefont {Wang}}, \bibinfo
  {author} {\bibfnamefont {H.}~\bibnamefont {Peelaers}},\ and\ \bibinfo
  {author} {\bibfnamefont {C.~G.}\ \bibnamefont {Van~de Walle}},\ }\href
  {https://doi.org/10.1063/5.0019915} {\bibfield  {journal} {\bibinfo
  {journal} {APL Mater.}\ }\textbf {\bibinfo {volume} {8}},\ \bibinfo {pages}
  {091105} (\bibinfo {year} {2020})}\BibitemShut {NoStop}%
\bibitem [{\citenamefont {Tsai}\ \emph {et~al.}(2010)\citenamefont {Tsai},
  \citenamefont {Bierwagen}, \citenamefont {White},\ and\ \citenamefont
  {Speck}}]{tsai2010beta}%
  \BibitemOpen
  \bibfield  {author} {\bibinfo {author} {\bibfnamefont {M.-Y.}\ \bibnamefont
  {Tsai}}, \bibinfo {author} {\bibfnamefont {O.}~\bibnamefont {Bierwagen}},
  \bibinfo {author} {\bibfnamefont {M.~E.}\ \bibnamefont {White}},\ and\
  \bibinfo {author} {\bibfnamefont {J.~S.}\ \bibnamefont {Speck}},\ }\href
  {https://doi.org/10.1116/1.3294715} {\bibfield  {journal} {\bibinfo
  {journal} {J. Vac. Sci. Technol., A}\ }\textbf {\bibinfo {volume} {28}},\
  \bibinfo {pages} {354} (\bibinfo {year} {2010})}\BibitemShut {NoStop}%
\bibitem [{\citenamefont {Kaun}\ \emph {et~al.}(2015)\citenamefont {Kaun},
  \citenamefont {Wu},\ and\ \citenamefont {Speck}}]{kaun2015beta}%
  \BibitemOpen
  \bibfield  {author} {\bibinfo {author} {\bibfnamefont {S.~W.}\ \bibnamefont
  {Kaun}}, \bibinfo {author} {\bibfnamefont {F.}~\bibnamefont {Wu}},\ and\
  \bibinfo {author} {\bibfnamefont {J.~S.}\ \bibnamefont {Speck}},\ }\href
  {https://doi.org/10.1116/1.4922340} {\bibfield  {journal} {\bibinfo
  {journal} {J. Vac. Sci. Technol., A}\ }\textbf {\bibinfo {volume} {33}},\
  \bibinfo {pages} {041508} (\bibinfo {year} {2015})}\BibitemShut {NoStop}%
\bibitem [{\citenamefont {Sasaki}\ \emph {et~al.}(2013)\citenamefont {Sasaki},
  \citenamefont {Higashiwaki}, \citenamefont {Kuramata}, \citenamefont
  {Masui},\ and\ \citenamefont {Yamakoshi}}]{2013_sasaki_MBE}%
  \BibitemOpen
  \bibfield  {author} {\bibinfo {author} {\bibfnamefont {K.}~\bibnamefont
  {Sasaki}}, \bibinfo {author} {\bibfnamefont {M.}~\bibnamefont {Higashiwaki}},
  \bibinfo {author} {\bibfnamefont {A.}~\bibnamefont {Kuramata}}, \bibinfo
  {author} {\bibfnamefont {T.}~\bibnamefont {Masui}},\ and\ \bibinfo {author}
  {\bibfnamefont {S.}~\bibnamefont {Yamakoshi}},\ }\href
  {https://doi.org/10.1016/j.jcrysgro.2013.02.015} {\bibfield  {journal}
  {\bibinfo  {journal} {J. Cryst. Growth}\ }\textbf {\bibinfo {volume} {378}},\
  \bibinfo {pages} {591} (\bibinfo {year} {2013})}\BibitemShut {NoStop}%
\bibitem [{\citenamefont {Okumura}\ \emph {et~al.}(2014)\citenamefont
  {Okumura}, \citenamefont {Kita}, \citenamefont {Sasaki}, \citenamefont
  {Kuramata}, \citenamefont {Higashiwaki},\ and\ \citenamefont
  {Speck}}]{2014_okumura_MBE}%
  \BibitemOpen
  \bibfield  {author} {\bibinfo {author} {\bibfnamefont {H.}~\bibnamefont
  {Okumura}}, \bibinfo {author} {\bibfnamefont {M.}~\bibnamefont {Kita}},
  \bibinfo {author} {\bibfnamefont {K.}~\bibnamefont {Sasaki}}, \bibinfo
  {author} {\bibfnamefont {A.}~\bibnamefont {Kuramata}}, \bibinfo {author}
  {\bibfnamefont {M.}~\bibnamefont {Higashiwaki}},\ and\ \bibinfo {author}
  {\bibfnamefont {J.~S.}\ \bibnamefont {Speck}},\ }\href
  {https://doi.org/10.7567/APEX.7.095501} {\bibfield  {journal} {\bibinfo
  {journal} {Appl. Phys. Express}\ }\textbf {\bibinfo {volume} {7}},\ \bibinfo
  {pages} {095501} (\bibinfo {year} {2014})}\BibitemShut {NoStop}%
\bibitem [{\citenamefont {Oshima}\ \emph {et~al.}(2018)\citenamefont {Oshima},
  \citenamefont {Ahmadi}, \citenamefont {Kaun}, \citenamefont {Wu},\ and\
  \citenamefont {Speck}}]{2018_oshima_MBE}%
  \BibitemOpen
  \bibfield  {author} {\bibinfo {author} {\bibfnamefont {Y.}~\bibnamefont
  {Oshima}}, \bibinfo {author} {\bibfnamefont {E.}~\bibnamefont {Ahmadi}},
  \bibinfo {author} {\bibfnamefont {S.}~\bibnamefont {Kaun}}, \bibinfo {author}
  {\bibfnamefont {F.}~\bibnamefont {Wu}},\ and\ \bibinfo {author}
  {\bibfnamefont {J.~S.}\ \bibnamefont {Speck}},\ }\href
  {https://doi.org/10.1088/1361-6641/aa9c4d} {\bibfield  {journal} {\bibinfo
  {journal} {Semicond. Sci. Technol.}\ }\textbf {\bibinfo {volume} {33}},\
  \bibinfo {pages} {015013} (\bibinfo {year} {2018})}\BibitemShut {NoStop}%
\bibitem [{\citenamefont {Azizie}\ \emph {et~al.}(2023)\citenamefont {Azizie},
  \citenamefont {Hensling}, \citenamefont {Gorsak}, \citenamefont {Kim},
  \citenamefont {Pieczulewski}, \citenamefont {Dryden}, \citenamefont
  {Senevirathna}, \citenamefont {Coye}, \citenamefont {Shang}, \citenamefont
  {Steele}, \citenamefont {Vogt}, \citenamefont {Parker}, \citenamefont
  {Birkhölzer}, \citenamefont {McCandless}, \citenamefont {Jena},
  \citenamefont {Xing}, \citenamefont {Liu}, \citenamefont {Williams},
  \citenamefont {Green}, \citenamefont {Chabak}, \citenamefont {Muller},
  \citenamefont {Neal}, \citenamefont {Mou}, \citenamefont {Thompson},
  \citenamefont {Nair},\ and\ \citenamefont {Schlom}}]{2023_Azizie_S-MBE}%
  \BibitemOpen
  \bibfield  {author} {\bibinfo {author} {\bibfnamefont {K.}~\bibnamefont
  {Azizie}}, \bibinfo {author} {\bibfnamefont {F.~V.~E.}\ \bibnamefont
  {Hensling}}, \bibinfo {author} {\bibfnamefont {C.~A.}\ \bibnamefont
  {Gorsak}}, \bibinfo {author} {\bibfnamefont {Y.}~\bibnamefont {Kim}},
  \bibinfo {author} {\bibfnamefont {N.~A.}\ \bibnamefont {Pieczulewski}},
  \bibinfo {author} {\bibfnamefont {D.~M.}\ \bibnamefont {Dryden}}, \bibinfo
  {author} {\bibfnamefont {M.~K.~I.}\ \bibnamefont {Senevirathna}}, \bibinfo
  {author} {\bibfnamefont {S.}~\bibnamefont {Coye}}, \bibinfo {author}
  {\bibfnamefont {S.-L.}\ \bibnamefont {Shang}}, \bibinfo {author}
  {\bibfnamefont {J.}~\bibnamefont {Steele}}, \bibinfo {author} {\bibfnamefont
  {P.}~\bibnamefont {Vogt}}, \bibinfo {author} {\bibfnamefont {N.~A.}\
  \bibnamefont {Parker}}, \bibinfo {author} {\bibfnamefont {Y.~A.}\
  \bibnamefont {Birkhölzer}}, \bibinfo {author} {\bibfnamefont {J.~P.}\
  \bibnamefont {McCandless}}, \bibinfo {author} {\bibfnamefont
  {D.}~\bibnamefont {Jena}}, \bibinfo {author} {\bibfnamefont {H.~G.}\
  \bibnamefont {Xing}}, \bibinfo {author} {\bibfnamefont {Z.-K.}\ \bibnamefont
  {Liu}}, \bibinfo {author} {\bibfnamefont {M.~D.}\ \bibnamefont {Williams}},
  \bibinfo {author} {\bibfnamefont {A.~J.}\ \bibnamefont {Green}}, \bibinfo
  {author} {\bibfnamefont {K.}~\bibnamefont {Chabak}}, \bibinfo {author}
  {\bibfnamefont {D.~A.}\ \bibnamefont {Muller}}, \bibinfo {author}
  {\bibfnamefont {A.~T.}\ \bibnamefont {Neal}}, \bibinfo {author}
  {\bibfnamefont {S.}~\bibnamefont {Mou}}, \bibinfo {author} {\bibfnamefont
  {M.~O.}\ \bibnamefont {Thompson}}, \bibinfo {author} {\bibfnamefont {H.~P.}\
  \bibnamefont {Nair}},\ and\ \bibinfo {author} {\bibfnamefont {D.~G.}\
  \bibnamefont {Schlom}},\ }\href {https://doi.org/10.1063/5.0139622}
  {\bibfield  {journal} {\bibinfo  {journal} {APL Mater.}\ }\textbf {\bibinfo
  {volume} {11}},\ \bibinfo {pages} {041102} (\bibinfo {year}
  {2023})}\BibitemShut {NoStop}%
\bibitem [{\citenamefont {Meng}\ \emph {et~al.}(2022)\citenamefont {Meng},
  \citenamefont {Feng}, \citenamefont {Bhuiyan},\ and\ \citenamefont
  {Zhao}}]{2022_MengLY_MOCVD}%
  \BibitemOpen
  \bibfield  {author} {\bibinfo {author} {\bibfnamefont {L.}~\bibnamefont
  {Meng}}, \bibinfo {author} {\bibfnamefont {Z.}~\bibnamefont {Feng}}, \bibinfo
  {author} {\bibfnamefont {A.~F. M. A.~U.}\ \bibnamefont {Bhuiyan}},\ and\
  \bibinfo {author} {\bibfnamefont {H.}~\bibnamefont {Zhao}},\ }\href
  {https://doi.org/10.1021/acs.cgd.2c00290} {\bibfield  {journal} {\bibinfo
  {journal} {Cryst. Growth Des.}\ }\textbf {\bibinfo {volume} {22}},\ \bibinfo
  {pages} {3896} (\bibinfo {year} {2022})},\ \bibinfo {note} {publisher:
  American Chemical Society}\BibitemShut {NoStop}%
\bibitem [{\citenamefont {Tang}\ \emph {et~al.}(2022)\citenamefont {Tang},
  \citenamefont {Ma}, \citenamefont {Zhang}, \citenamefont {Zhou},
  \citenamefont {Zhang}, \citenamefont {Zhang}, \citenamefont {Chen},
  \citenamefont {Wei}, \citenamefont {Lin}, \citenamefont {Mudiyanselage},
  \citenamefont {Fu},\ and\ \citenamefont {Zhang}}]{tang2022high}%
  \BibitemOpen
  \bibfield  {author} {\bibinfo {author} {\bibfnamefont {W.}~\bibnamefont
  {Tang}}, \bibinfo {author} {\bibfnamefont {Y.}~\bibnamefont {Ma}}, \bibinfo
  {author} {\bibfnamefont {X.}~\bibnamefont {Zhang}}, \bibinfo {author}
  {\bibfnamefont {X.}~\bibnamefont {Zhou}}, \bibinfo {author} {\bibfnamefont
  {L.}~\bibnamefont {Zhang}}, \bibinfo {author} {\bibfnamefont
  {X.}~\bibnamefont {Zhang}}, \bibinfo {author} {\bibfnamefont
  {T.}~\bibnamefont {Chen}}, \bibinfo {author} {\bibfnamefont {X.}~\bibnamefont
  {Wei}}, \bibinfo {author} {\bibfnamefont {W.}~\bibnamefont {Lin}}, \bibinfo
  {author} {\bibfnamefont {D.~H.}\ \bibnamefont {Mudiyanselage}}, \bibinfo
  {author} {\bibfnamefont {H.}~\bibnamefont {Fu}},\ and\ \bibinfo {author}
  {\bibfnamefont {B.}~\bibnamefont {Zhang}},\ }\href
  {https://doi.org/10.1063/5.0092754} {\bibfield  {journal} {\bibinfo
  {journal} {Appl. Phys. Lett.}\ }\textbf {\bibinfo {volume} {120}},\ \bibinfo
  {pages} {212103} (\bibinfo {year} {2022})}\BibitemShut {NoStop}%
\bibitem [{\citenamefont {Shan}\ \emph {et~al.}(2005)\citenamefont {Shan},
  \citenamefont {Liu}, \citenamefont {Lee}, \citenamefont {Lee}, \citenamefont
  {Kim},\ and\ \citenamefont {Shin}}]{2005_Shan_ALD}%
  \BibitemOpen
  \bibfield  {author} {\bibinfo {author} {\bibfnamefont {F.~K.}\ \bibnamefont
  {Shan}}, \bibinfo {author} {\bibfnamefont {G.~X.}\ \bibnamefont {Liu}},
  \bibinfo {author} {\bibfnamefont {W.~J.}\ \bibnamefont {Lee}}, \bibinfo
  {author} {\bibfnamefont {G.~H.}\ \bibnamefont {Lee}}, \bibinfo {author}
  {\bibfnamefont {I.~S.}\ \bibnamefont {Kim}},\ and\ \bibinfo {author}
  {\bibfnamefont {B.~C.}\ \bibnamefont {Shin}},\ }\href
  {https://doi.org/10.1063/1.1980535} {\bibfield  {journal} {\bibinfo
  {journal} {J. Appl. Phys. (Melville, NY, U. S.)}\ }\textbf {\bibinfo {volume}
  {98}},\ \bibinfo {pages} {023504} (\bibinfo {year} {2005})}\BibitemShut
  {NoStop}%
\bibitem [{\citenamefont {Mauze}\ \emph {et~al.}(2020)\citenamefont {Mauze},
  \citenamefont {Zhang}, \citenamefont {Itoh}, \citenamefont {Wu},\ and\
  \citenamefont {Speck}}]{2020_mauze_MOCATAXY}%
  \BibitemOpen
  \bibfield  {author} {\bibinfo {author} {\bibfnamefont {A.}~\bibnamefont
  {Mauze}}, \bibinfo {author} {\bibfnamefont {Y.}~\bibnamefont {Zhang}},
  \bibinfo {author} {\bibfnamefont {T.}~\bibnamefont {Itoh}}, \bibinfo {author}
  {\bibfnamefont {F.}~\bibnamefont {Wu}},\ and\ \bibinfo {author}
  {\bibfnamefont {J.~S.}\ \bibnamefont {Speck}},\ }\href
  {https://doi.org/10.1063/1.5135930} {\bibfield  {journal} {\bibinfo
  {journal} {APL Mater.}\ }\textbf {\bibinfo {volume} {8}},\ \bibinfo {pages}
  {021104} (\bibinfo {year} {2020})}\BibitemShut {NoStop}%
\bibitem [{\citenamefont {Wagner}\ \emph {et~al.}(2014)\citenamefont {Wagner},
  \citenamefont {Baldini}, \citenamefont {Gogova}, \citenamefont {Schmidbauer},
  \citenamefont {Schewski}, \citenamefont {Albrecht}, \citenamefont {Galazka},
  \citenamefont {Klimm},\ and\ \citenamefont
  {Fornari}}]{2014_wagner_homoepitaxial_stacking}%
  \BibitemOpen
  \bibfield  {author} {\bibinfo {author} {\bibfnamefont {G.}~\bibnamefont
  {Wagner}}, \bibinfo {author} {\bibfnamefont {M.}~\bibnamefont {Baldini}},
  \bibinfo {author} {\bibfnamefont {D.}~\bibnamefont {Gogova}}, \bibinfo
  {author} {\bibfnamefont {M.}~\bibnamefont {Schmidbauer}}, \bibinfo {author}
  {\bibfnamefont {R.}~\bibnamefont {Schewski}}, \bibinfo {author}
  {\bibfnamefont {M.}~\bibnamefont {Albrecht}}, \bibinfo {author}
  {\bibfnamefont {Z.}~\bibnamefont {Galazka}}, \bibinfo {author} {\bibfnamefont
  {D.}~\bibnamefont {Klimm}},\ and\ \bibinfo {author} {\bibfnamefont
  {R.}~\bibnamefont {Fornari}},\ }\href
  {https://doi.org/10.1002/pssa.201330092} {\bibfield  {journal} {\bibinfo
  {journal} {Phys. Status Solidi A}\ }\textbf {\bibinfo {volume} {211}},\
  \bibinfo {pages} {27} (\bibinfo {year} {2014})}\BibitemShut {NoStop}%
\bibitem [{\citenamefont {Fiedler}\ \emph {et~al.}(2017)\citenamefont
  {Fiedler}, \citenamefont {Schewski}, \citenamefont {Baldini}, \citenamefont
  {Galazka}, \citenamefont {Wagner}, \citenamefont {Albrecht},\ and\
  \citenamefont {Irmscher}}]{2017_fiedler_influence_stackingfault}%
  \BibitemOpen
  \bibfield  {author} {\bibinfo {author} {\bibfnamefont {A.}~\bibnamefont
  {Fiedler}}, \bibinfo {author} {\bibfnamefont {R.}~\bibnamefont {Schewski}},
  \bibinfo {author} {\bibfnamefont {M.}~\bibnamefont {Baldini}}, \bibinfo
  {author} {\bibfnamefont {Z.}~\bibnamefont {Galazka}}, \bibinfo {author}
  {\bibfnamefont {G.}~\bibnamefont {Wagner}}, \bibinfo {author} {\bibfnamefont
  {M.}~\bibnamefont {Albrecht}},\ and\ \bibinfo {author} {\bibfnamefont
  {K.}~\bibnamefont {Irmscher}},\ }\href {https://doi.org/10.1063/1.4993748}
  {\bibfield  {journal} {\bibinfo  {journal} {J. Appl. Phys.}\ }\textbf
  {\bibinfo {volume} {122}},\ \bibinfo {pages} {165701} (\bibinfo {year}
  {2017})}\BibitemShut {NoStop}%
\bibitem [{\citenamefont {Eisner}\ \emph {et~al.}(2020)\citenamefont {Eisner},
  \citenamefont {Ranga}, \citenamefont {Bhattacharyya}, \citenamefont
  {Krishnamoorthy},\ and\ \citenamefont {Scarpulla}}]{2020_eisner_201_010}%
  \BibitemOpen
  \bibfield  {author} {\bibinfo {author} {\bibfnamefont {B.~A.}\ \bibnamefont
  {Eisner}}, \bibinfo {author} {\bibfnamefont {P.}~\bibnamefont {Ranga}},
  \bibinfo {author} {\bibfnamefont {A.}~\bibnamefont {Bhattacharyya}}, \bibinfo
  {author} {\bibfnamefont {S.}~\bibnamefont {Krishnamoorthy}},\ and\ \bibinfo
  {author} {\bibfnamefont {M.~A.}\ \bibnamefont {Scarpulla}},\ }\href
  {https://doi.org/10.1063/5.0022043} {\bibfield  {journal} {\bibinfo
  {journal} {J. Appl. Phys.}\ }\textbf {\bibinfo {volume} {128}},\ \bibinfo
  {pages} {195703} (\bibinfo {year} {2020})}\BibitemShut {NoStop}%
\bibitem [{\citenamefont {Sdoeung}\ \emph {et~al.}(2021)\citenamefont
  {Sdoeung}, \citenamefont {Sasaki}, \citenamefont {Masuya}, \citenamefont
  {Kawasaki}, \citenamefont {Hirabayashi}, \citenamefont {Kuramata},\ and\
  \citenamefont {Kasu}}]{2021_sdoeung_stacking_leakage}%
  \BibitemOpen
  \bibfield  {author} {\bibinfo {author} {\bibfnamefont {S.}~\bibnamefont
  {Sdoeung}}, \bibinfo {author} {\bibfnamefont {K.}~\bibnamefont {Sasaki}},
  \bibinfo {author} {\bibfnamefont {S.}~\bibnamefont {Masuya}}, \bibinfo
  {author} {\bibfnamefont {K.}~\bibnamefont {Kawasaki}}, \bibinfo {author}
  {\bibfnamefont {J.}~\bibnamefont {Hirabayashi}}, \bibinfo {author}
  {\bibfnamefont {A.}~\bibnamefont {Kuramata}},\ and\ \bibinfo {author}
  {\bibfnamefont {M.}~\bibnamefont {Kasu}},\ }\href
  {https://doi.org/10.1063/5.0049761} {\bibfield  {journal} {\bibinfo
  {journal} {Appl. Phys. Lett.}\ }\textbf {\bibinfo {volume} {118}},\ \bibinfo
  {pages} {172106} (\bibinfo {year} {2021})}\BibitemShut {NoStop}%
\bibitem [{\citenamefont {Bermudez}(2006)}]{bermudez2006the}%
  \BibitemOpen
  \bibfield  {author} {\bibinfo {author} {\bibfnamefont {V.}~\bibnamefont
  {Bermudez}},\ }\href {https://doi.org/10.1016/j.chemphys.2005.08.051}
  {\bibfield  {journal} {\bibinfo  {journal} {Chem. Phys.}\ }\textbf {\bibinfo
  {volume} {323}},\ \bibinfo {pages} {193} (\bibinfo {year}
  {2006})}\BibitemShut {NoStop}%
\bibitem [{\citenamefont {Peelaers}\ and\ \citenamefont {Van~de
  Walle}(2015)}]{peelaers2015brillouin}%
  \BibitemOpen
  \bibfield  {author} {\bibinfo {author} {\bibfnamefont {H.}~\bibnamefont
  {Peelaers}}\ and\ \bibinfo {author} {\bibfnamefont {C.~G.}\ \bibnamefont
  {Van~de Walle}},\ }\href {https://doi.org/10.1002/pssb.201451551} {\bibfield
  {journal} {\bibinfo  {journal} {Phys. Status Solidi B}\ }\textbf {\bibinfo
  {volume} {252}},\ \bibinfo {pages} {828} (\bibinfo {year}
  {2015})}\BibitemShut {NoStop}%
\bibitem [{\citenamefont {Wang}\ \emph {et~al.}(2023)\citenamefont {Wang},
  \citenamefont {Mu}, \citenamefont {Speck},\ and\ \citenamefont {Van~de
  Walle}}]{wang2023first}%
  \BibitemOpen
  \bibfield  {author} {\bibinfo {author} {\bibfnamefont {M.}~\bibnamefont
  {Wang}}, \bibinfo {author} {\bibfnamefont {S.}~\bibnamefont {Mu}}, \bibinfo
  {author} {\bibfnamefont {J.~S.}\ \bibnamefont {Speck}},\ and\ \bibinfo
  {author} {\bibfnamefont {C.~G.}\ \bibnamefont {Van~de Walle}},\ }\href
  {https://doi.org/https://doi.org/10.1002/admi.202300318} {\bibfield
  {journal} {\bibinfo  {journal} {Adv. Mater. Interfaces}\ }\textbf {\bibinfo
  {volume} {n/a}},\ \bibinfo {pages} {2300318} (\bibinfo {year}
  {2023})}\BibitemShut {NoStop}%
\bibitem [{\citenamefont {Zhao}\ \emph {et~al.}(2023)\citenamefont {Zhao},
  \citenamefont {Byggm{\"a}star}, \citenamefont {He}, \citenamefont {Nordlund},
  \citenamefont {Djurabekova},\ and\ \citenamefont {Hua}}]{SFzhao2023complex}%
  \BibitemOpen
  \bibfield  {author} {\bibinfo {author} {\bibfnamefont {J.}~\bibnamefont
  {Zhao}}, \bibinfo {author} {\bibfnamefont {J.}~\bibnamefont
  {Byggm{\"a}star}}, \bibinfo {author} {\bibfnamefont {H.}~\bibnamefont {He}},
  \bibinfo {author} {\bibfnamefont {K.}~\bibnamefont {Nordlund}}, \bibinfo
  {author} {\bibfnamefont {F.}~\bibnamefont {Djurabekova}},\ and\ \bibinfo
  {author} {\bibfnamefont {M.}~\bibnamefont {Hua}},\ }\href
  {https://doi.org/10.1038/s41524-023-01117-1} {\bibfield  {journal} {\bibinfo
  {journal} {npj Comput. Mater.}\ }\textbf {\bibinfo {volume} {9}},\ \bibinfo
  {pages} {159} (\bibinfo {year} {2023})}\BibitemShut {NoStop}%
\bibitem [{\citenamefont {Zhao}\ \emph {et~al.}(2016)\citenamefont {Zhao},
  \citenamefont {Baibuz}, \citenamefont {Vernieres}, \citenamefont
  {Grammatikopoulos}, \citenamefont {Jansson}, \citenamefont {Nagel},
  \citenamefont {Steinhauer}, \citenamefont {Sowwan}, \citenamefont {Kuronen},
  \citenamefont {Nordlund},\ and\ \citenamefont
  {Djurabekova}}]{SFzhao2016formation}%
  \BibitemOpen
  \bibfield  {author} {\bibinfo {author} {\bibfnamefont {J.}~\bibnamefont
  {Zhao}}, \bibinfo {author} {\bibfnamefont {E.}~\bibnamefont {Baibuz}},
  \bibinfo {author} {\bibfnamefont {J.}~\bibnamefont {Vernieres}}, \bibinfo
  {author} {\bibfnamefont {P.}~\bibnamefont {Grammatikopoulos}}, \bibinfo
  {author} {\bibfnamefont {V.}~\bibnamefont {Jansson}}, \bibinfo {author}
  {\bibfnamefont {M.}~\bibnamefont {Nagel}}, \bibinfo {author} {\bibfnamefont
  {S.}~\bibnamefont {Steinhauer}}, \bibinfo {author} {\bibfnamefont
  {M.}~\bibnamefont {Sowwan}}, \bibinfo {author} {\bibfnamefont
  {A.}~\bibnamefont {Kuronen}}, \bibinfo {author} {\bibfnamefont
  {K.}~\bibnamefont {Nordlund}},\ and\ \bibinfo {author} {\bibfnamefont
  {F.}~\bibnamefont {Djurabekova}},\ }\href
  {https://doi.org/10.1021/acsnano.6b01024} {\bibfield  {journal} {\bibinfo
  {journal} {ACS Nano}\ }\textbf {\bibinfo {volume} {10}},\ \bibinfo {pages}
  {4684} (\bibinfo {year} {2016})}\BibitemShut {NoStop}%
\bibitem [{\citenamefont {Zhao}\ \emph {et~al.}(2017)\citenamefont {Zhao},
  \citenamefont {Cao}, \citenamefont {Palmer}, \citenamefont {Nordlund},\ and\
  \citenamefont {Djurabekova}}]{SFzhao2017formation}%
  \BibitemOpen
  \bibfield  {author} {\bibinfo {author} {\bibfnamefont {J.}~\bibnamefont
  {Zhao}}, \bibinfo {author} {\bibfnamefont {L.}~\bibnamefont {Cao}}, \bibinfo
  {author} {\bibfnamefont {R.~E.}\ \bibnamefont {Palmer}}, \bibinfo {author}
  {\bibfnamefont {K.}~\bibnamefont {Nordlund}},\ and\ \bibinfo {author}
  {\bibfnamefont {F.}~\bibnamefont {Djurabekova}},\ }\href
  {https://doi.org/10.1103/PhysRevMaterials.1.066002} {\bibfield  {journal}
  {\bibinfo  {journal} {Phys. Rev. Mater.}\ }\textbf {\bibinfo {volume} {1}},\
  \bibinfo {pages} {066002} (\bibinfo {year} {2017})}\BibitemShut {NoStop}%
\bibitem [{\citenamefont {Plimpton}(1995)}]{lammps1995}%
  \BibitemOpen
  \bibfield  {author} {\bibinfo {author} {\bibfnamefont {S.}~\bibnamefont
  {Plimpton}},\ }\href {https://doi.org/10.1006/jcph.1995.1039} {\bibfield
  {journal} {\bibinfo  {journal} {J. Comp. Phys.}\ }\textbf {\bibinfo {volume}
  {117}},\ \bibinfo {pages} {1} (\bibinfo {year} {1995})}\BibitemShut {NoStop}%
\bibitem [{\citenamefont {Stukowski}(2010)}]{2010_ovito_stukowski}%
  \BibitemOpen
  \bibfield  {author} {\bibinfo {author} {\bibfnamefont {A.}~\bibnamefont
  {Stukowski}},\ }\href {https://doi.org/10.1088/0965-0393/18/1/015012}
  {\bibfield  {journal} {\bibinfo  {journal} {Modelling Simul. Mater. Sci.
  Eng.}\ }\textbf {\bibinfo {volume} {18}},\ \bibinfo {pages} {015012}
  (\bibinfo {year} {2010})}\BibitemShut {NoStop}%
\bibitem [{\citenamefont {Hammond}(2020)}]{2020W-zmethod}%
  \BibitemOpen
  \bibfield  {author} {\bibinfo {author} {\bibfnamefont {K.~D.}\ \bibnamefont
  {Hammond}},\ }\href {https://doi.org/10.1016/j.cpc.2019.106862} {\bibfield
  {journal} {\bibinfo  {journal} {Comput. Phys. Commun.}\ }\textbf {\bibinfo
  {volume} {247}},\ \bibinfo {pages} {106862} (\bibinfo {year}
  {2020})}\BibitemShut {NoStop}%
\bibitem [{\citenamefont {Schewski}\ \emph {et~al.}(2016)\citenamefont
  {Schewski}, \citenamefont {Baldini}, \citenamefont {Irmscher}, \citenamefont
  {Fiedler}, \citenamefont {Markurt}, \citenamefont {Neuschulz}, \citenamefont
  {Remmele}, \citenamefont {Schulz}, \citenamefont {Wagner}, \citenamefont
  {Galazka},\ and\ \citenamefont
  {Albrecht}}]{2016_schewski_evolution_stacking_001}%
  \BibitemOpen
  \bibfield  {author} {\bibinfo {author} {\bibfnamefont {R.}~\bibnamefont
  {Schewski}}, \bibinfo {author} {\bibfnamefont {M.}~\bibnamefont {Baldini}},
  \bibinfo {author} {\bibfnamefont {K.}~\bibnamefont {Irmscher}}, \bibinfo
  {author} {\bibfnamefont {A.}~\bibnamefont {Fiedler}}, \bibinfo {author}
  {\bibfnamefont {T.}~\bibnamefont {Markurt}}, \bibinfo {author} {\bibfnamefont
  {B.}~\bibnamefont {Neuschulz}}, \bibinfo {author} {\bibfnamefont
  {T.}~\bibnamefont {Remmele}}, \bibinfo {author} {\bibfnamefont
  {T.}~\bibnamefont {Schulz}}, \bibinfo {author} {\bibfnamefont
  {G.}~\bibnamefont {Wagner}}, \bibinfo {author} {\bibfnamefont
  {Z.}~\bibnamefont {Galazka}},\ and\ \bibinfo {author} {\bibfnamefont
  {M.}~\bibnamefont {Albrecht}},\ }\href {https://doi.org/10.1063/1.4971957}
  {\bibfield  {journal} {\bibinfo  {journal} {J. Appl. Phys.}\ }\textbf
  {\bibinfo {volume} {120}},\ \bibinfo {pages} {225308} (\bibinfo {year}
  {2016})}\BibitemShut {NoStop}%
\bibitem [{\citenamefont {Yamaguchi}\ and\ \citenamefont
  {Kuramata}(2018)}]{2018_yamaguchi_stacking}%
  \BibitemOpen
  \bibfield  {author} {\bibinfo {author} {\bibfnamefont {H.}~\bibnamefont
  {Yamaguchi}}\ and\ \bibinfo {author} {\bibfnamefont {A.}~\bibnamefont
  {Kuramata}},\ }\href {https://doi.org/10.1107/S1600576718011093} {\bibfield
  {journal} {\bibinfo  {journal} {J. Appl. Crystallogr.}\ }\textbf {\bibinfo
  {volume} {51}},\ \bibinfo {pages} {1372} (\bibinfo {year}
  {2018})}\BibitemShut {NoStop}%
\end{thebibliography}%

\end{document}